\documentclass{emulateapj}
\usepackage{lscape}

\newcommand{\cm}{\ensuremath{\rm{cm^{-2}}}}
\newcommand{\kms}{\ensuremath{\,\rm{km\,s^{-1}}}}
\newcommand{\etal}{et al.\ }

\newcommand{\Msun}{\ensuremath{M_{\odot}}}
\newcommand{\Vrot}{\ensuremath{V_{rot}}}
\newcommand{\ropt}{\ensuremath{r_{opt}}}
\newcommand{\ints}{\ensuremath{\int{I}\mathrm{d} v}}
\newcommand{\wsini}{\ensuremath{W\sin{i}}}
\newcommand{\magasec}{\ensuremath{\rm{mag} \, \rm{asec}^{-2}}}
\newcommand{\Iband}{$I$-band}
\newcommand{\ha}{{H}{$\alpha$}}
\newcommand{\hi}{\ion{H}{1}}

\newcommand{\hs}{\hspace*{10pt}}

\newcommand{\lcdm}{$\Lambda$CDM}

\newcommand{\none}{NGC~1324}
\newcommand{\utwo}{UGC~2849}
\newcommand{\eso}{ESO~563G21}
\newcommand{\ufiveo}{NGC~2862}
\newcommand{\ufiveone}{NGC~2955}
\newcommand{\ueight}{IC~4202}
\newcommand{\uten}{NGC~6195}
\newcommand{\uone}{UGC~11455}

\defcitealias{h99}{H99}
\defcitealias{cat05}{C05}
\defcitealias{cvg}{CvG91}

\slugcomment{To appear in the Astronomical Journal}

\shorttitle{Structure of Rapidly Rotating Spirals: I}
\shortauthors{Spekkens \& Giovanelli}

\begin{document}


\title{The Structure of Rapidly Rotating Late-Type Spiral Galaxies: I. Photometry, \hi\ and Optical Kinematics}


\author{Kristine Spekkens\altaffilmark{1} \& Riccardo Giovanelli\altaffilmark{2}}
\affil{Center for Radiophysics and Space Research, Cornell University, Ithaca, NY 14853}


\altaffiltext{1}{Present address: Jansky Fellow, National Radio Astronomy Observatory; and Department of Physics and Astronomy, Rutgers, the State University of New Jersey, 136 Frelinghuysen Road, Piscataway, NJ 08854; spekkens@physics.rutgers.edu} 
\altaffiltext{2}{National Astronomy and Ionosphere Center (NAIC), Cornell University, Ithaca, NY 14853; riccardo@astro.cornell.edu. NAIC is operated by Cornell University under cooperative agreement with the National Science Foundation (NSF).}


\begin{abstract}
We present \Iband\ photometry, long-slit optical spectroscopy, and new aperture synthesis \hi\ observations for eight late-type spiral galaxies with rotation velocities in the range $243\kms \lesssim V_{rot} \lesssim 308\kms$. The sample will be used to study the structure and angular momentum of disks at the high-mass end of the spiral galaxy population; here we discuss the basic properties of these ``fast rotators'', and derive hybrid optical/\hi\ rotation curves for each. Despite the presence of \hi\ warps and low-mass companions in many systems, their kinematics are regular and there is excellent agreement between optical and \hi\ tracers near the optical radius \ropt.  At high inclinations at which projection effects are negligible, the sample galaxies exhibit flat, featureless rotation curves out to their last measured points at $1.7-3.5\,\ropt$. The intermediate-inclination systems are also consistent with a constant rotation amplitude for $r \gtrsim0.5\,\ropt$. We therefore find no evidence for declining rotation curves at the high-mass end of the late-type spiral galaxy population. Combining our data with the compilation of spiral galaxies with reliable outer \hi\ kinematics from the work of Casertano and van Gorkom, we find no convincing trends between logarithmic outer rotation curve slopes and rotation amplitudes or surface brightnesses for galaxies with $\Vrot \gtrsim 220\,\kms$.  Correlations between these slopes and morphological types or disk scale lengths are also marginal in this regime. 

\end{abstract}


\section{Introduction}
\label{fast1:intro}

\begin{deluxetable*}{ccccccc}
\tablecaption{Basic Properties of the Fast Rotator Sample \label{fast1_tab:basic}} 
\tablewidth{0pt}
\tablehead{ \colhead{Galaxy} & \colhead{$\alpha_c$} & \colhead{$\delta_c$} & \multicolumn{2}{c}{type} & \colhead{$V_{\odot}$} & \colhead{$D$}\\
 & \colhead{(J2000)} & \colhead{(J2000)} & \multicolumn{2}{c}{\hs} & \colhead{(\kms)} & \colhead{(Mpc)}\\
\colhead{(1)} & \colhead{(2)} & \colhead{(3)} & \multicolumn{2}{c}{(4)} & \colhead{(5)} & \colhead{(6)}
}
\startdata
\none & 03 25 01.6 & -05 44 42 & Sb: & (3) & 5672  & 79.4\\
\utwo & 03 45 32.9 & +44 51 18 & Scd:& (6) & 8142 & 116.0\\
\eso &  08 47 16.4 & -20 02 09 & SAbc: sp & (4) & 4581 & 70.1 \\
\ufiveo & 09 24 55.3 & +26 46 29 & Sb & (3)& 4100 & 63.0\\
\ufiveone & 09 41 16.6 & +35 52 56 & (R')SA(r)b & (3) & 7008 & 105.0\\
\ueight & 13 08 31.5 & +24 42 01 & Sbc & (4) & 7127 & 103.9\\
\uten & 16 36 32.6 & +39 01 37 & Sb & (3) & 9022 & 130.5\\
\uone & 19 29 56.0 & +72 06 44 & Sc & (6) & 5393 & 76.1\\
\enddata
\tablecomments{Col. (1): NGC/IC designation where available, otherwise UGC/ESO number. Cols. (2)-(3): Galaxy coordinates (J2000). Col.(4): Morphological
type, as given in the NASA/IPAC Extragalactic Database. Corresponding RC3 type numbers are given in parentheses \citep{rc3}. Col. (5): Heliocentric (optical) radial velocity, derived from archived single-dish spectra as in \citet{springob05}. Col. (7): Distance corresponding to $V_{\odot}$, computed using $\alpha_c$, $\delta_c$ and $H_0=70\,\kms \, \rm{Mpc}^{-1}$.}
\end{deluxetable*}

The dynamics of rapidly rotating spiral galaxies provide insight into the structure of systems at the high-mass extreme of the galaxy population. Their properties are essential for understanding well-established trends between the rotation velocity $V_{rot}$ of a disk galaxy and its other basic parameters, such as morphology, luminosity, surface brightness and baryon fraction. Observations of massive nearby galaxies also constitute a benchmark for studies of more distant systems, since the most luminous counterparts to local spiral galaxies are more easily detected at high redshift. ``Fast rotators'' in the local Universe thus hold important clues to the formation and evolution of the spiral galaxy population as a whole.

 However, most recent studies of the kinematics of disk galaxies have focused on low mass and/or low surface brightness late-type systems (e.g. de Blok et al. 1996; de Blok \& McGaugh 1997; Swaters 1999; Swaters et al. 2000; McGaugh et al. 2001; Marchesini et al. 2002;  van den Bosch \& Swaters 2001;  van den Bosch et al. 2001; de Blok et al. 2001, 2003; Swaters et al. 2003; Rhee et al. 2004; Gentile et al. 2005; Simon et al. 2005; Spekkens et al. 2005; Valenzuela et al. 2005), while others have examined intermediate-mass systems \citep[e.g.][hereafter G04]{gar02,kregel04a,kregel05,gentile04}. Observations of early type disk galaxies that span a wide range of morphological types and luminosities are now underway (Noordermeer et al. 2005), but high-quality, homogeneously processed data for late-type massive galaxies have so far been lacking. To fill this gap we present photometry and kinematics for eight of these systems, which will be used to model the dark and luminous contributions to their mass profiles in a forthcoming paper (Spekkens \& Giovanelli 2006, in preparation).

It has long been recognized that both the shapes and amplitudes of spiral galaxy rotation curves (RCs) change systematically with luminosity (e.g. Roberts \& Rots 1973; Kyazumov 1984; Burstein \& Rubin 1985). Average RC shapes within the optical radii \ropt\ of late-type spiral galaxies have been studied extensively by Persic and collaborators (Persic \& Salucci 1988, 1990a,b, 1991; Persic et al. 1996), using long-slit optical spectra. In their ``universal rotation curve'' (URC) model, systems with absolute \Iband\ luminosities $M_I^o\lesssim-22.0$ ($V_{rot}\gtrsim 200\,\kms$) have RCs that peak at $\sim0.8\ropt$ and decline at larger galactocentric radii $r$. However, the URC parametrization tends to underpredict the outer RC amplitudes of high surface brightness systems (e.g. Courteau 1997).  Average optical RC shapes for a much larger sample of spiral galaxies have since been compiled by Catinella et al. (2006), who find no evidence for declining RCs within \ropt\ in even the most luminous systems.

 Neutral hydrogen (\hi) is an effective tracer of late-type spiral galaxy kinematics beyond \ropt, since the \hi\ disk is typically $50\%$ larger than the stellar one (e.g. Broeils \& Rhee 1997). The outer RC shapes derived from \ion{H}{1} observations are typically more uncertain than those in the optical, however, because measurements of the inclination $i$ are required to correct observed velocities for viewing geometry. While $i$ is relatively constant across the optical disks of spiral galaxies and can be robustly measured from photometry, warps are common in the \ion{H}{1} layer beyond \ropt\ \citep[e.g.][]{gar02} and variations in $i$ in these regions can be difficult to constrain with available data (see, e.g. \S\ref{fast1:inti}).  Nonetheless, the majority of spiral galaxies have \hi\ RCs that are relatively flat in their outer parts (e.g. Bosma 1978, 1981a,b), although the possibility that many RCs become Keplerian within $r\sim3\ropt$ cannot be ruled out (Honma \& Sofue 1997). There are also some unambiguous detections of decreasing RCs beyond \ropt\ (see the compilation by Honma \& Sofue 1997). Two such cases are NGC~2683 and NGC~3521, where large ($\sim50\,\kms$, or $0.25\,V_{rot}$) declines are detected between 1 and 3 \ropt\ (Casertano \& van Gorkom 1991, hereafter CvG91). \citetalias{cvg} also find correlations between the logarithmic RC slopes beyond 2/3\ropt\  and the rotation amplitudes, scale-lengths, surface brightnesses, and morphological types for a diverse sample of late-type spiral galaxies with reliable \hi\ RCs. These trends suggest that the RCs of massive systems -- compact, high surface brightness ones in particular --  decrease in amplitude or decline steadily beyond \ropt. 

 Observationally, spiral galaxies with $V_{rot} \gtrsim 350\, \kms$ are rare, particularly for later morphological types.  While $V_{rot}$ for the fastest known rotator in the local Universe exceeds $500\,\kms$ (UGC~12591; Giovanelli et al. 1986), the distribution of \hi\ spectral line widths in large samples of spiral galaxies truncates abruptly near $W \sim 2V_{rot} \sim 550\,\kms$ (e.g. Koribalski et al. 2004; Springob et al. 2005).  Elliptical galaxies exhibit a similar upper limit to their central velocity dispersions, which can be understood in the context of the $\Lambda$CDM structure formation paradigm provided that the stars in their cores formed prior to $z\sim6$ and behave like collisionless particles thereafter (Loeb \& Peebles 2003). There is no analogous explanation for the upper bound to \Vrot\ in spiral galaxies.  One possibility is that the progenitors of spiral galaxies with $V_{rot} > 350\,\kms$ seldom form long-lived disks. In particular, if the latter are stabilized against bar formation by massive halos (Ostriker \& Peebles 1973, Efstathiou et al. 1982; Christodoulou et al. 1995a,b) and their baryonic components scale with parent halo masses and angular momenta (e.g. Dalcanton et al. 1997; Mo et al. 1998; de Jong \& Lacey 2000), then a stability criterion of the form $\lambda \gtrsim m_d$ arises, where $\lambda$ is the halo spin parameter and $m_d$ is the cold baryon fraction (Mo et al. 1998). Since collisionless halos with $\lambda > 0.1$ are rare (e.g. Bullock et al. 2001), massive systems that retain most of their baryons during galaxy formation would seldom have large enough $\lambda$ to form stable disks. In this scenario, massive spiral galaxies have large $\lambda$. The angular momentum distributions of local fast rotators probe the stability of disks in high-mass systems, and thus this explanation for the observed upper limit to $V_{rot}$.

 We have obtained \hi\ aperture synthesis observations for eight late-type, rapidly rotating spiral galaxies in the local Universe, each with \Iband\ photometry and optical spectroscopy archived in the SFI++ database maintained at Cornell University. We combine the optical and \hi\ kinematics to produce hybrid RCs for each system, that have arcsecond resolution within \ropt\ and extend beyond the stellar disk into the region in which dark matter dominates the potential. In a future paper we will use these data to constrain the dark and luminous contributions to the derived kinematics and to compute the angular momentum distributions and $\lambda$ for a wide range of baryon mass-to-light ratios (K. Spekkens \& R. Giovanelli 2006, in preparation). Here we present our observations and concentrate on deriving accurate optical+\hi\ RCs that will underpin our models. We also compare the outer RC shapes obtained to those of the systems compiled by \citetalias{cvg} and reevaluate the evidence for correlations between these shapes and other galaxy properties with the fast rotator and \citetalias{cvg} samples combined. 

The organization of this paper is as follows. We describe the sample selection and present available SFI++ catalog photometry and optical kinematics in \S\ref{fast1:sselec}. We present the aperture synthesis \hi\ observations and the related analysis in \S\ref{fast1:HI} and describe the derivation of hybrid optical/\hi\ RCs in \S\ref{fast1:kinboth}. We summarize our results and discuss the outer RC shapes of the fast rotator sample in \S\ref{fast1:sample}. Notes on individual systems are given in Appendix~\ref{fast1:ind}.

 Unless otherwise stated, we convert angular sizes to physical scales by adopting $H_0=70\,\kms\ \, \rm{Mpc}^{-1}$ and give radial velocities following the heliocentric, optical definition.

\section{Sample Selection and SFI++ Data}
\label{fast1:sselec}

 The fast rotator sample is selected from the SFI++ database maintained at Cornell University. The SFI++ is a compilation of \Iband\ photometry and \Vrot\ measurements for several thousand late-type spiral galaxies, optimized for studies of the large-scale mass distribution in the local Universe. The catalog includes several data sets published by our group (\citealp[][hereafter H99]{h99}; \citealp[][and references therein]{dale00}; \citealp{vogt04,springob05}; \citealp[][hereafter C05]{cat05}) as well as newly acquired data, and is fully described elsewhere (\citealp{karenthesis,christhesis}; C. M. Springob et al. 2006, in preparation). 

 Sample candidates have both single-dish \hi\ profiles and long-slit optical spectroscopy archived in the SFI++. Of these, we select eight galaxies with the largest \Vrot\ that (1) show no evidence for interactions, close companions or disturbed optical morphologies, and (2) have \hi\ line strengths and optical sizes suitable for \hi\ aperture synthesis observations with the Very Large Array\footnote{The Very Large Array is a facility of the National Radio Astronomy Observatory (NRAO). The NRAO is a facility of the NSF, operated under cooperative agreement with Associated Universities, Inc.} (VLA). The basic properties of the fast rotator sample are presented in Table~\ref{fast1_tab:basic}. 

 The mean rotational velocity for the sample galaxies is $\Vrot\ \sim 275\,\kms$; they are therefore at the high-mass end of the local spiral galaxy population. Unlike the compilation of \citet{saglia88}, however, we find no evidence for discrepancies between the system properties and the velocity, size, magnitude, and surface brightness scaling relations for the SFI++ \citep{mythesis}. 
Moreover, despite their large absolute scales both the global gas fractions and surface densities of the galaxies are within the ranges expected for normal spiral galaxies of the same morphological types (see \S\ref{fast1:global}). The fast rotator sample thus represents an extreme of the normal spiral galaxy population rather than a separate class of object.
 

\begin{deluxetable*}{cccccccccccc}
\tablecaption{\Iband\ Photometry \label{fast1_tab:iband}}
\tabletypesize{\scriptsize}
\tablewidth{0pt}
\tablehead{ \colhead{NGC/IC} & \colhead{$M_I^o$} & \colhead{$r_d^o$} & \colhead{$\mu_{e,I}$} & \colhead{$B_T^0-I$} & \colhead{$r_{opt}$}  & \colhead{$r_{235}^o$} & \colhead{$\theta_d$} & \colhead{$eps_d$} & \colhead{$i_d$} & \colhead{$B/D_{min}$} & \colhead{$B/D_{max}$} \\
 & \colhead{mag} & \colhead{(kpc)} & \colhead{(mag$\,$asec$^{-2}$)} & \colhead{(mag)} &  \colhead{(kpc)} & \colhead{(kpc)} & \colhead{($^{\circ}$)} & & \colhead{($^{\circ}$)} & & \\
\colhead{(1)} & \colhead{(2)} & \colhead{(3)} & \colhead{(4)} & \colhead{(5)} & \colhead{(6)} & \colhead{(7)} & \colhead{(8)} & \colhead{(9)} & \colhead{(10)} & \colhead{(11)} & \colhead{(12)}
}
\startdata
\none & -23.6 & 5.6 & 18.7 & \nodata & 19.7& 27.2 &$139.9 \pm 0.7$ & $0.72 \pm 0.04$ & 78 & 0.02 & 0.08 \\
\utwo & -23.5 & 6.2 & 19.0 & \nodata & 21.0 & 20.5  &$82 \pm 2$ & $0.55 \pm 0.2$ & 66 & 0.02 & 0.04 \\
ESO 563G21 & -23.6 & 6.9 & 19.1 & 1.36 & 25.7 & 28.1 &$165.9 \pm 0.5$ & $0.77 \pm 0.02$ & 83 & 0.06 & 0.06 \\
\ufiveo & -23.2 & 4.4 & 18.6 & 1.77 &  15.0 & 22.6 & $114.4 \pm 0.9$ & $0.79 \pm 0.01$ & 86 & 0.0 & 0.03 \\
\ufiveone & -23.3 & 4.9 & 18.7 & \nodata & 20.2 & 23.5 & $159 \pm 5$ & $0.42 \pm 0.06$ & 56 & 0.3 & 0.3 \\
\ueight & -23.1 & 4.5 & 18.8 & 1.95 & 16.7 & 21.2 & $142.0 \pm 0.1$ & $0.82 \pm 0.01$ & 90 & 0.01 & 0.1 \\
\uten & -23.7 & 6.1 & 18.8 & 1.76 & 23.0 & 26.2 &$39 \pm 5$ & $0.5 \pm 0.1$ & 62 & 0.03 & 0.6 \\
\uone & -23.3 & 5.8 & 19.1 & 2.12 & 21.0 & 23.7 & $62 \pm 1$ & $0.83 \pm 0.02$ & 90 & 0.0 & 0.03 \\
\enddata
\tablecomments{Col. (2): Absolute \Iband\ magnitude computed using $D$ from Table~\ref{fast1_tab:basic}, corrected for internal and galactic extinction. Col.(3): 
Exponential disk scale-length, corrected to a face-on perspective. Col. (4): \Iband\ effective central surface brightness, computed using $M_I^o$ from col. (2) and $r_d^o$ from col. (3). Col. (5): $B-I$ color estimated using $B_T^0$ from the RC3 \citep{rc3} and $M_I^o$ from col. (2).   Col. (6): Optical radius, within which 83\% of the 
total \Iband\ light is contained. Col. (7): Radius of the 23.5$\,\magasec$ isophote, corrected to a face-on perspective. Col. (8): Average position angle in the disk-fitting region. Col (9): Average ellipticity ($eps=1-b/a$, where $a$ and $b$ are major and minor axes) in the disk-fitting region. Col. (10): Average inclination in the disk-fitting region, computed from col. (9) assuming an intrinsic axial ratio $q=0.2$. Col. (11): ``Minimum'' ratio of bulge light to disk light, from decomposition with the exponential disk of col. (2). Col. (12):  ``Maximum'' ratio of bulge light to disk light, from decomposition with a disk fit in the ``dip'' of the surface brightness profile (see text).}
\end{deluxetable*}

\subsection{\Iband\ Images and Photometry}
\label{fast1:iband}

\begin{figure}
\epsscale{1.2}
\plotone{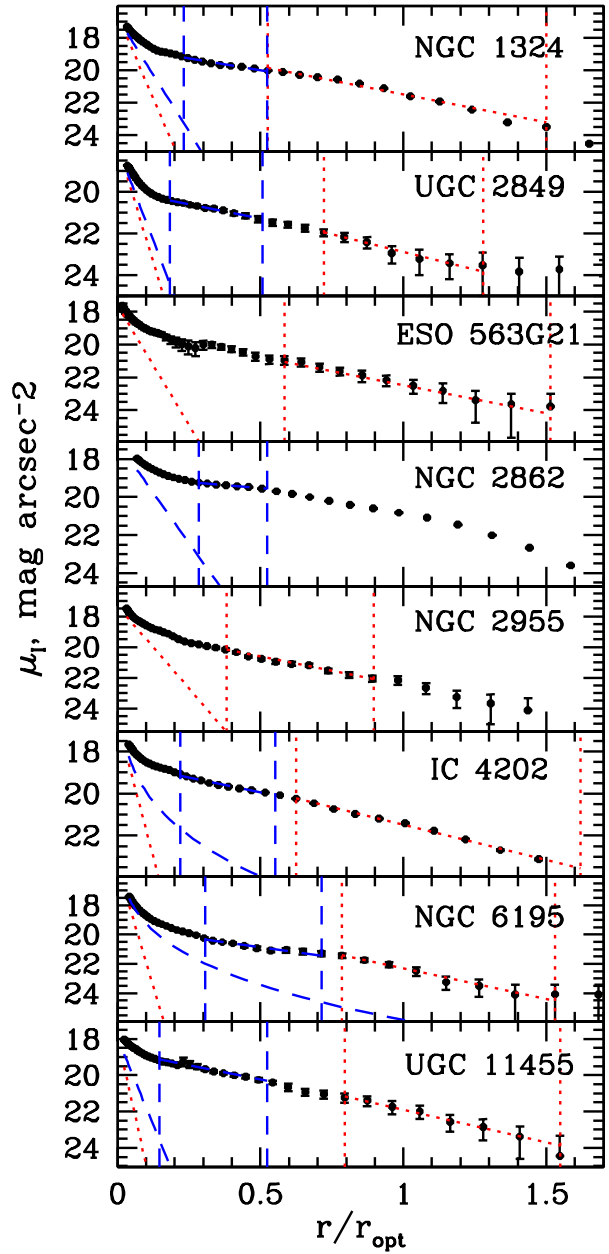} 
\caption{\Iband\ surface brightness profiles for the fast rotator sample. In each panel, points between the vertical dotted lines constrain the exponential disk in the ``minimal'' bulge computation, while points between the dashed vertical lines constrain the exponential disk in the ``maximal'' bulge computation. In each case, the disk fits are given by the lines through the points, and the resulting minimal and maximal bulge fits are represented by the dotted and dashed diagonal lines, respectively. See the electronic edition of the Journal for a color version of this figure.}
\label{fast1_fig:sb}
\end{figure}

 The \Iband\ images and photometry for the sample are mined from the SFI++, and a thorough description of the data acquisition and reduction techniques can be found in \citetalias{h99}. Below, we summarize this procedure and describe our bulge-disk decomposition technique in some detail.


 The \Iband\ images were obtained during a series of observing runs by our group between 1988 and 1997, using the 1.3m McGraw-Hill telescope at the MDM Observatory\footnote{The Observatory is owned and operated by a consortium of five universities: the University of Michigan, Dartmouth College, the Ohio State University, Columbia University, and Ohio University.} as well as the 0.9m telescopes at the Kitt Peak National Observatory and the Cerro Tololo Inter-American Observatory\footnote{The Kitt Peak National Observatory and Cerro-Tololo Inter-American Observatory are part of the National Optical Astronomy Observatory (NOAO), operated by Associated Universities for Research in Astronomy (AURA), Inc., under a cooperative agreement with the NSF.}. The images were homogeneously calibrated using standard procedures within the IRAF STSDAS environment\footnote{IRAF is distributed by NOAO, which is operated by AURA under cooperative agreement with the NSF. STSDAS is distributed by the Space Telescope Science Institute, which is operated by AURA under contract with the National Aeronautics and Space Administration.} in the manner described in \citetalias{h99}. 

 Following calibration, the resulting image analysis was performed by M. P. Haynes for all galaxies. Photometric quantities were extracted by fitting concentric isophotal ellipses to the galaxy emission after masking nearby stars. The resulting surface brightness profiles are shown in Fig.~\ref{fast1_fig:sb}. We define the optical radius $r_{opt}$ as the radius within which 83\% of the \Iband\ profile light is contained \citep{pss96}; \ropt\ is equivalent to 3.2 scale lengths in a pure exponential disk. The radius $r_{235}$ of the 23.5$\,\magasec$ isophote is also measured. The average disk ellipticity $\mathrm{eps}_d$ and position angle $\theta_d$ are determined by averaging measurements\footnote{For the highly inclined systems $\mathrm{eps}_d$ is taken as the maximum value in this region instead of the mean, since the isophote-fitting method artificially decreases the latter \citepalias{h99}.} for ellipses in a region of the outer profile likely dominated by disk emission, as described by \citet{gio94}. The disk scale length $r_d$ is then derived from the slope of a linear fit to points in this region. The total apparent magnitude $m_I$ is measured at $8r_d$, extrapolated to this level from the disk fit if required. The disk inclination $i_d$ is then computed from $\mathrm{eps}_d$ assuming an intrinsic axial ratio $q=0.2$, and the photometric quantities are corrected to a face-on perspective ($r_{235}^o$, $r_d^o$, $m_I^o$) using the prescriptions of \citet{gio95}.  We then compute the effective central surface brightness $\mu_{e,I}$ following \citetalias{cvg}, defined as the central surface brightness of a pure exponential disk with total luminosity $m_I^o$ and scale length $r_d^o$. The bulge contributions to the total \Iband\ light are small for the fast rotators (see below), and thus $\mu_{e,I}$ approaches the extrapolated disk central surface brightness tabulated by \citetalias{h99}. The derived photometric quantities for each system are given in Table~\ref{fast1_tab:iband}.

 The surface brightness profiles are then decomposed into disk and bulge contributions. To estimate the bulge component, we first subtract an exponential disk from the profile.  The remaining light is then fitted with both spherical exponential and de Vaucouleurs (1948) bulges, and the best-fitting profile of the two is adopted as the bulge distribution. This bulge profile is then subtracted from the total surface brightness distribution, and the remaining light is assigned to the disk. The decomposition therefore preserves both the total \Iband\ flux as well as the small-scale features of the profiles, which are assumed to stem from the disk.

 The fraction of \Iband\ flux assigned to the bulge clearly depends on the exponential disk first subtracted from the surface brightness distribution. However, the best-fitting disk determined from one region of the surface brightness distribution may deviate significantly at other $r$ in the profile due to irregularities in the baryon distribution or to extinction effects. This is the case for many of the fast rotators, which show a ``dip'' in their surface brightness distributions for $0.1\ropt\ \lesssim r \lesssim 0.4\ropt$ (Fig.~\ref{fast1_fig:sb}). To estimate the range of reasonable bulge contributions for each galaxy, we therefore carry out the procedure described above starting from two distinct exponential disks: the fit from H99 yields a suitable ``minimal'' bulge contribution, while one estimated in the region of the profile dip produces a suitable ``maximal'' bulge contribution. Because of extinction effects in the surface brightness profiles, the actual bulge component is likely bracketed by these extremes.

 The resulting bulge profiles are shown in Fig.~\ref{fast1_fig:sb}. For each galaxy, points between the vertical dotted lines are used to estimate the exponential disk in the minimal bulge computation \citepalias{h99}, while those between the dashed vertical lines produce the exponential disk in the maximal bulge estimate. The corresponding exponential disks are shown by the dotted and dashed lines through the points, and the resulting bulge fits are shown by the dotted and dashed diagonal lines at small $r$. Note that for \ufiveo\ only the maximal bulge fit is plausible, while for \eso\ and \ufiveone\ the maximal and minimal estimates are identical; for these systems, only one bulge fit is therefore displayed. The resulting minimal and maximal ratios of bulge light to disk light are given in the last two columns of Table~\ref{fast1_tab:iband}. As expected for late-type systems, the bulge component is dwarfed by the disk component even in the ``maximum'' bulge scenario. The bulge is thus not expected to contribute substantially to the observed kinematics, and it is likely reasonable to model the fast rotators as pure disk systems by assigning all of the \Iband\ light to a thin axisymmetric component.  The impact of these different decompositions on the inferred structure of each system will be fully explored at the mass modeling stage (K. Spekkens \& R. Giovanelli 2006, in preparation).

 For five of the eight fast rotators, unmasked \Iband\ images are archived in the SFI++ in addition to the photometry. For these systems, we determine a plate scale for each image by matching 15-20 stars in the vicinity of the galaxy to those in analogous Digitized Sky Survey\footnote{The DSS was produced at the Space Telescope Science Institute under US Government grant NAG W-2166. The images of these surveys are based on photographic data obtained using the Oschin Schmidt Telescope on Palomar Mountain and the UK Schmidt Telescope. The plates were processed into the present compressed digital form with the permission of these institutions.} (DSS) images and by solving for the pixel size, center and rotation angle. The positional accuracy of this linear solution is judged to be better than $1''$. The available \Iband\ images are shown in gray scale in Figs~\ref{fast1_fig:1324summ}a, \ref{fast1_fig:5010summ}a, and \ref{fast1_fig:8220summ}a -- \ref{fast1_fig:11455summ}a. 

 For clarity in the text that follows we adopt $\theta$ and $i$ as abbreviations for position angle and inclination, respectively, while $\theta_d$ and $i_d$ specify the values from \Iband\ photometry that are listed in cols. (8) and (10) of Table~\ref{fast1_tab:iband}.

\subsection{Long-Slit Optical Spectroscopy}
\label{fast1:halpha}

 The optical spectra for the fast rotator sample have also been mined from the SFI++. The data reduction and analysis are detailed in \citetalias{cat05}; here we summarize the procedure.

 Long-slit optical spectra for seven of the eight fast rotators were obtained by our group between 1990 and 2002, using the Double Spectrograph on the 200 inch (5 m)Hale Telescope at Palomar Observatory\footnote{The Hale Telescope, Palomar Observatory, is operated under a collaborative agreement between the California Institute of Technology, its divisions Caltech Optical Observatories and the Jet Propulsion Laboratory (operated for NASA), and Cornell University.}. The observing setup yielded a spectral resolution of $0.65\,$\AA$\,$pixel$^{-1}$ and a spatial resolution of $0.'' 468\,$pixel$^{-1}$ across a $2''\,\times\,125''$ slit. The long axis of the slit was placed at $\theta=\phi_l$ (Table~\ref{fast1_tab:kin}) for each system. In general $\phi_l = \theta_d$ from the \Iband\ images, within the uncertainties in the latter. The seeing full width at half-maximum (FWHM) was typically $1.5''$. The data were reduced and rotational velocities extracted using both standard IRAF packages and our own software. Where possible, the kinematics from the H$\alpha$ and \ion{N}{2} lines were combined to form a single optical RC \citepalias{cat05}. The RCs for \ufiveo, \ufiveone, \ueight\ and \uten\ presented here have been published by Vogt et al. (2004).

 For \eso, we use the H$\alpha$ RC derived by \citet{mat92} obtained with a similar observing setup and reduction technique.  

 After calibration, all of the RCs were analyzed in the manner detailed in \citetalias{cat05}. Briefly, we fold each RC about the position determined from the best fitting URC to the points \citep{pss96}, and then fit a polyex function,
\begin{equation}
V_{pe}(r) = V_o \sin{i}(1-e^{-r/r_{pe}})(1+\beta r/r_{pe})\,\,,
\label{fast1:a2}
\end{equation}
 where $V_o \sin{i}$ sets the amplitude of the fit, $r_{pe}$ governs the inner RC slope, and $\beta$ determines the outer RC slope. Values of $r_{pe}$ for the fast rotators are given in Table~\ref{fast1_tab:kin}. Since \Vrot\ and the outer RC slope are better characterized by the hybrid RCs than the optical spectra alone, we do not tabulate the SFI++ $V_o$ and $\beta$ here. 

 Tables~\ref{fast1_tab:iband} and \ref{fast1_tab:kin} demonstrate that the sample galaxies fall into two groups, depending on $i_d$ and $r_{pe}$: \none, \utwo, \ufiveone, and \uten\ have $i_d\leq78^{\circ}$ and $1.1 \le r_{pe} \le 2.3\,$kpc, whereas \eso, \ufiveo, \ueight\ and \uone\ have $i_d\geq83^{\circ}$ and $r_{pe} > 4\,$kpc. The large $r_{pe}$ in the latter galaxies likely result from extinction effects, which become more pronounced with increasing $i$ \citep[e.g.][]{gio02}. The fast rotators thus divide into an intermediate $i$ sub-sample and a highly inclined sub-sample.

 The optical RCs for the fast rotator sample, corrected for projection effects using $i_d$ and resampled to a resolution of approximately $2''$ (3.2\arcsec\ for ES0~563G21) to reflect the seeing conditions, are shown by the triangles in Figs.~\ref{fast1_fig:edgeons}--\ref{fast1_fig:intinc}. The error bars on each point reflect either the accuracy with which the emission centroid could be determined in the optical spectrum, or an imposed minimum of 2/$\sin{i_d}\,\kms$.

\section{Aperture Synthesis \hi\ Observations}
\label{fast1:HI}

 The sample galaxies were observed with the VLA in its C configuration in separate runs during 2001 July -- August and 2002 October -- December, except for ESO~536G21 which was observed in BC configuration in 2002 October to prevent excessive antenna shadowing at low latitudes.  During each run the galaxy was targeted in 25--35 minute intervals, separated by 3--5 minute phase calibrator observations. The flux calibrator, also used as a bandpass calibrator, was observed at both the start and the end of each run. The raw on-source integration time for each system is listed in Table~\ref{fast1_tab:HIobs}.

The sample galaxies were at the pointing center of their corresponding runs, resulting in sky coverage within a primary beam FWHM=31.3\arcmin\ centered on each source.  
To adequately sample the full width of the integrated \hi\ line for the fast rotators, two 3.125 MHz bandpasses were staggered about their single-dish profile centers. This yielded contiguous frequency coverage across each source in $97\,$kHz spectral channels, as well as sufficient baseline to subtract continuum emission. The bandpasses overlapped near the band center, with the number of duplicate channels determined from the net bandwidth required for each observation.
 
 The data were calibrated using the Astronomical Image Processing System\footnote{{\tt http://www.aoc.nrao.edu/aips/}} (AIPS; Napier \etal 1993). For each run, standard editing and flux, phase, and bandpass calibration routines were applied separately to the two polarizations of each bandpass (Rupen 1999 and references therein). The bandpasses were then joined and the data in the overlapping channels averaged to produce a single band; the number of overlapping channels at the band center for each system is given in Table~\ref{fast1_tab:HIobs}. In most cases the continuum emission could be subtracted via linear fits to the real and imaginary Fourier components in the line-free frequency channels (e.g. van Langevelde \& Cotton 1990). For \none, \ufiveone\ and \ueight, models of strong continuum sources far from the pointing center were first subtracted (Ekers \& van Gorkom 1984; van Gorkom \& Ekers 1989). 

 Following the data reduction, strong radio frequency interference (RFI) was discovered in one or two channels near $1400\,$MHz in the \eso\ and \ufiveo\ data cubes. This well-known VLA birdie is caused by correlated noise from the seventh harmonic of the $200\,$MHz L-band local oscillator in the shortest east-west baselines, where the fringe rate is small (Bagri 1996). We excise this RFI following the method of Spekkens \etal (2004).  We remove baselines with east-west projections less than $25$ and $17\,$m in all spectral channels and clip contaminated channels above $0.7$ and $1.0\,$Jy for \eso\ and \ufiveo, respectively. In all, the excision flags less than 5\% of the available baselines. We checked for possible biases from this editing by examining residual data sets in the same manner as described in the Appendix of Spekkens \etal (2004), and found no evidence for any systematic flux loss.
 We measure an increased rms map noise $\sigma$ of 10\% at 4389\kms\ in the \eso\ data cube as a result of the clipping but find no change in the RFI-excised channels of the \ufiveo\ data cube.

 After calibration the data were imaged using both natural and uniform weighting schemes to emphasize \hi\ structures on different spatial scales.  The emission in each cube was then CLEANed (Clark 1980) of sidelobe contamination and corrected for the attenuation of the primary beam.  The synthesized beam parameters and the average rms noise $\bar{\sigma}$ per channel for the calibrated data sets are given in Table~\ref{fast1_tab:HIobs}. In general, the map noise $\sigma$ is $\sim10$\% lower than $\bar{\sigma}$ in the overlap channels. This gain in sensitivity is less than the factor $\sqrt{2}$ anticipated from averaging two independent channels together because of correlated noise from continuum subtraction and decreased sensitivity near the bandpass edge.

\subsection{\hi\ Data Presentation}
\label{fast1:analysis}

\begin{figure}
\epsscale{1.2}
\plotone{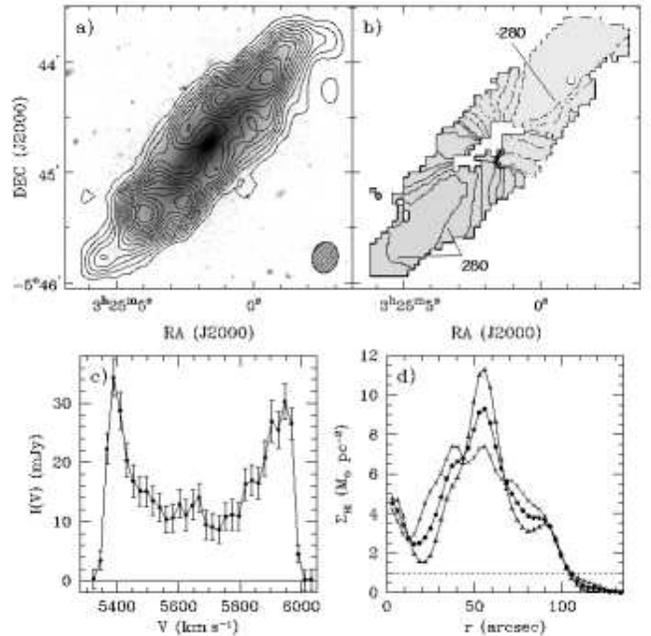}
\caption{\hi\ properties of \none. $(a)$ Uniformly weighted \hi\ total intensity contours overplotted on an \Iband\ image. Contours range from $3.8 \times 10^{20}$ to $3.0 \times 10^{21}\,\cm$ in $2.0 \times 10^{20}\,\cm$ intervals. The synthesized beam is shown in the bottom right corner. $(b)$ MET velocity field (shaded region).  Contours are plotted in 40\kms\ increments relative to $V_{sys}$ in Table~\ref{fast1_tab:properties}, with solid lines on the receding side and dotted lines on the approaching side. Some contours are labelled in kilometers per second for clarity.  $(c)$ Global profile. Error bars reflect 1$\sigma$ uncertainty contributions from $\bar{\sigma}$ over the emission region and a 5\% calibration error. $(d)$ Radial surface density profile $\Sigma_{HI}(r)$ corrected to a face-on perspective for the approaching side (crosses), the receding side (triangles), and their average (points). The dotted line shows $1\Msun\,\rm{pc} ^{-2}$.}
 \label{fast1_fig:1324summ}
\end{figure}

\begin{figure}
\epsscale{1.2}
\plotone{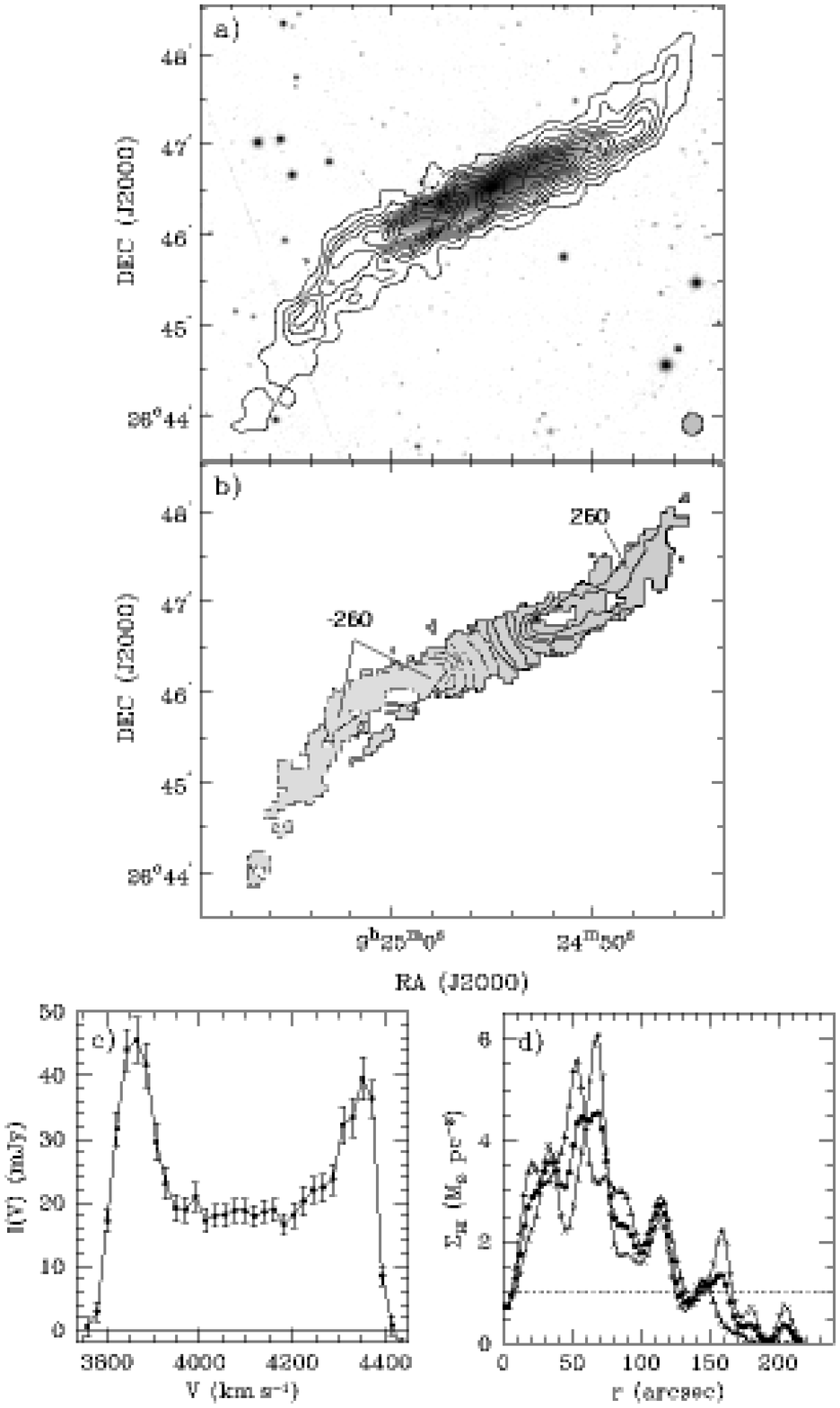}
\caption{Same as Fig.~\ref{fast1_fig:1324summ}, but for \ufiveo.  In $(a)$, \hi\ contours range from $1.3 \times 10^{20}$ to $3.8 \times 10^{21}\,\cm$  in $3.2 \times 10^{20}\,\cm$ intervals, overplotted on an \Iband\ image.}
 \label{fast1_fig:5010summ}
\end{figure}

\begin{figure}
\epsscale{1.2}
\plotone{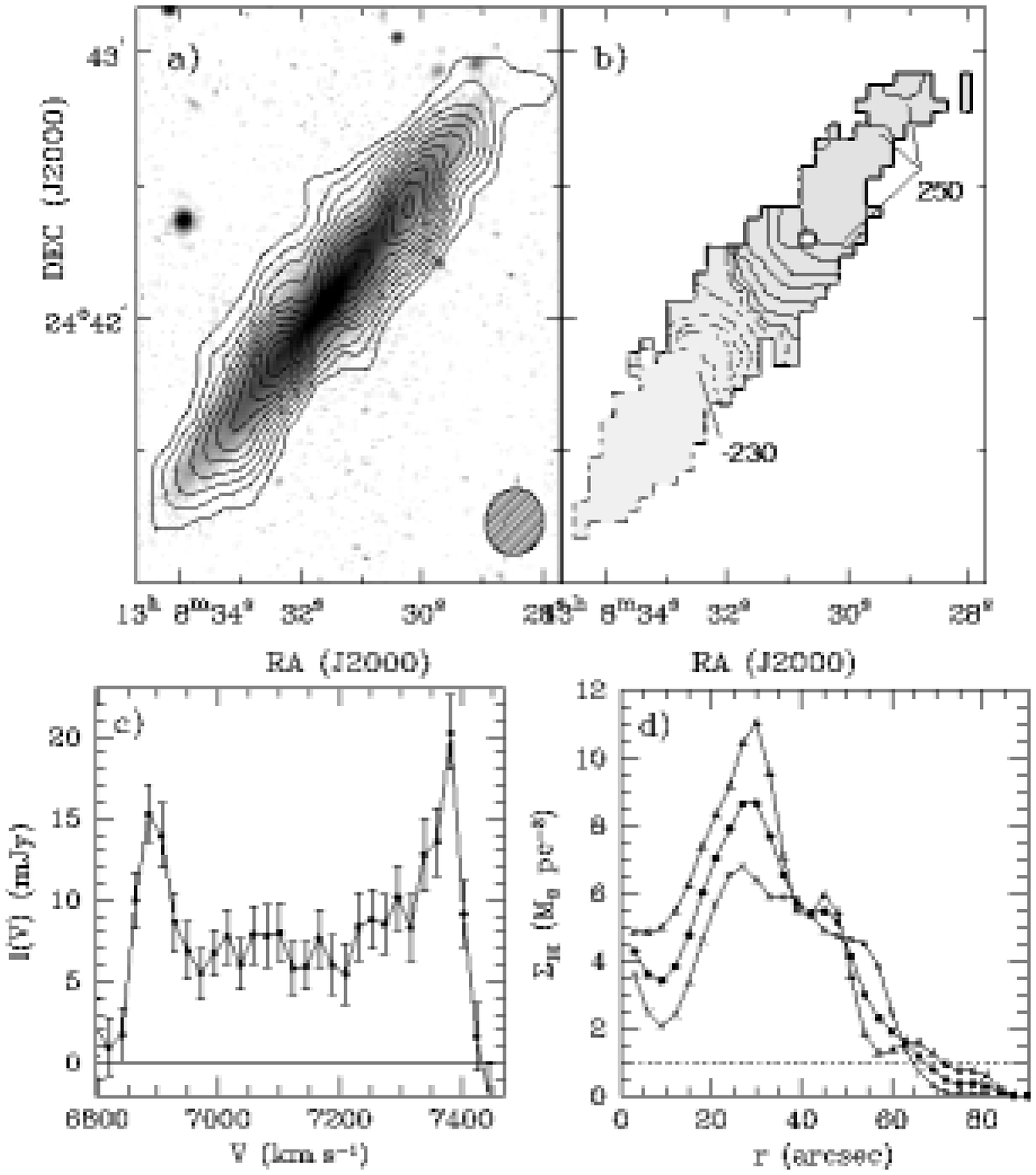}
\caption{Same as Fig.~\ref{fast1_fig:1324summ}, but for \ueight. In $(a)$, \hi\ contours range from $4.0 \times 10^{20}$ to $4.0 \times 10^{21}\,\cm$ in $3.3 \times 10^{20}\,\cm$ intervals, overplotted on an \Iband\ image.}
 \label{fast1_fig:8220summ}
\end{figure}

\begin{figure}
\epsscale{1.2}
\plotone{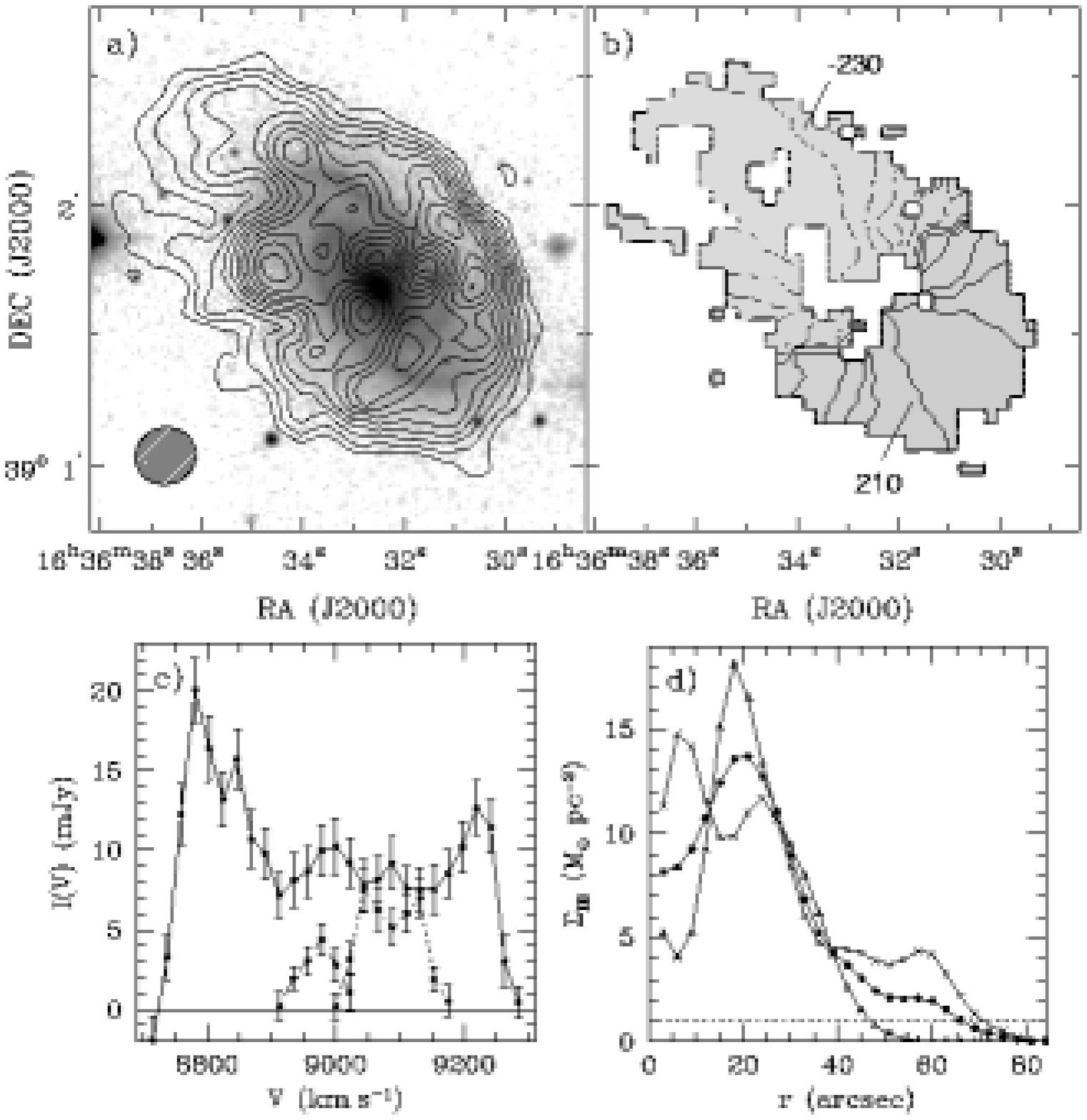}
\caption{Same as Fig.~\ref{fast1_fig:1324summ}, but for \uten. In $(a)$, \hi\ contours range from $2.7 \times 10^{20}$ to $4.8 \times 10^{21}\,\cm$ in $1.4 \times 10^{20}\,\cm$ intervals, overplotted on an \Iband\ image.}
 \label{fast1_fig:10469summ}
\end{figure}

\begin{figure}
\epsscale{1.2}
\plotone{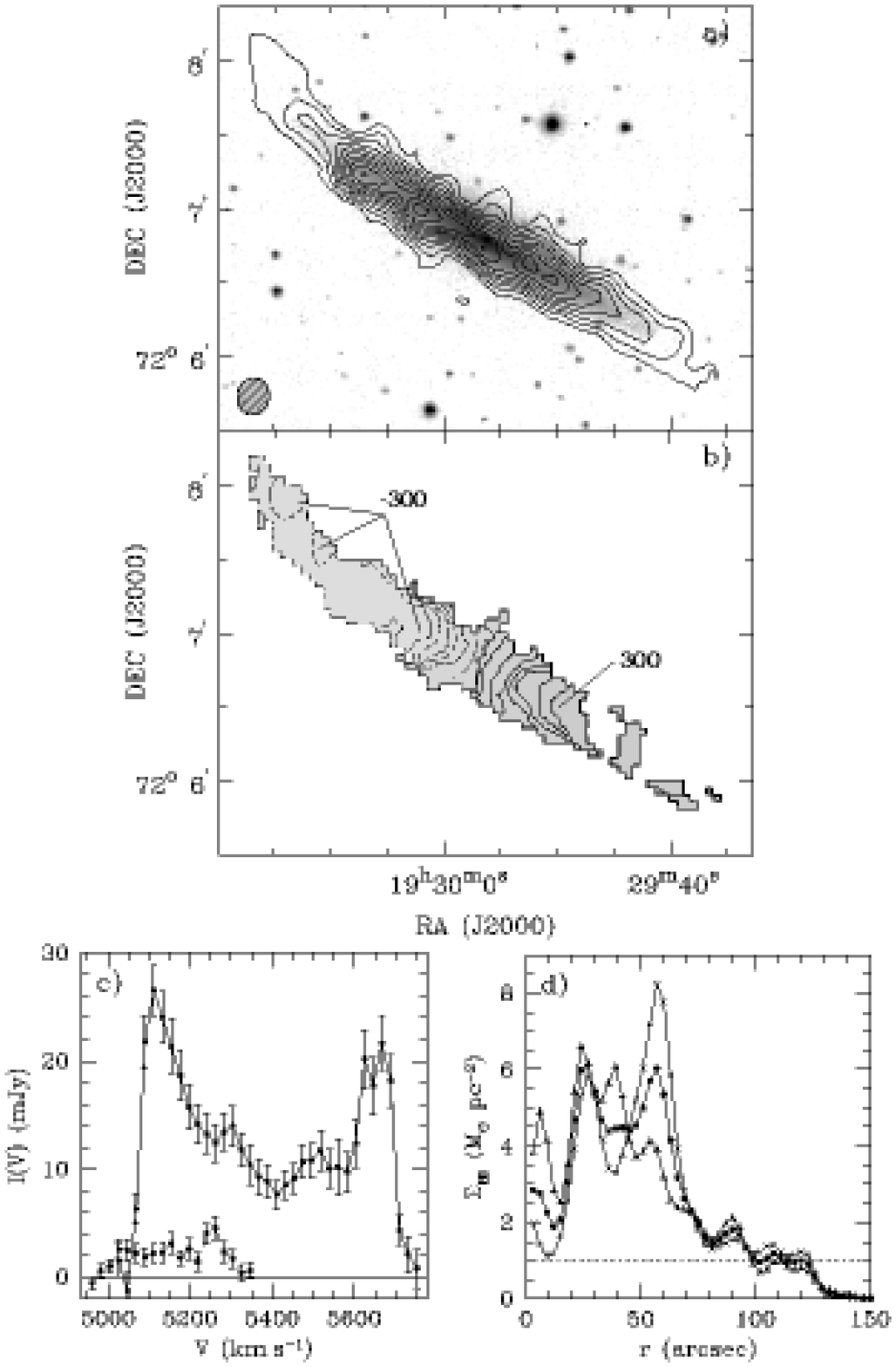}
\caption{Same as Fig.~\ref{fast1_fig:1324summ}, but for \uone. In $(a)$, \hi\ contours range from $3.9 \times 10^{20}$ to $4.5 \times 10^{21}\,\cm$ in $3.9 \times 10^{20}\,\cm$ intervals, overplotted on an \Iband\ image.}
 \label{fast1_fig:11455summ}
\end{figure}

 The \hi\ data for the fast rotators are presented in Figs~\ref{fast1_fig:1324summ}--\ref{fast1_fig:11455summ}, and their properties are summarized in Table~\ref{fast1_tab:properties}.   Channel maps are in Figure Set 13, and notes for individual systems are in Appendix~\ref{fast1:ind}. 

 In the interest of deriving high-quality kinematics in the outer \hi-rich parts of the disks, we trade sensitivity for angular resolution and analyze the uniformly weighted maps for all the fast rotators except \eso, where the signal-to-noise ratio (S/N) is too low with uniform weighting. We also perform a cursory search for companions in the naturally weighted cubes: we make detections in the vicinity of half the sample galaxies, and half again are previously uncataloged. The properties of the detected companions are given in Table~\ref{fast1_tab:propertiesc}. 
Throughout, we use a combination of AIPS and Groningen Image Processing System\footnote{{\tt http://www.astro.rug.nl/$\sim$gipsy/}} routines to manipulate and display the data.
 
\begin{deluxetable}{ccccc}
\tablecaption{Optical and \hi\ Kinematics \label{fast1_tab:kin}}
\tablewidth{0pt}
\tablehead{\colhead{NGC/IC} & \colhead{$\phi_l$} & \colhead{$r_{pe}$} & \colhead{\Vrot} & \colhead{$S$} \\
                            &  \colhead{($^{\circ}$)} & \colhead{(kpc)}  & \colhead{(\kms)} &         \\
              \colhead{(1)} & \colhead{(2)}    & \colhead{(3)}      & \colhead{(4)}   & \colhead{(5)}
}
\startdata
\none & 142 & 1.4 & $297 \pm 4$ & -0.10 \\
\utwo & 72 & 2.3 & $270 \pm 2$ & -0.04 \\
\eso & 164 & 4.1 & $308 \pm 2$ & 0.00\\
\ufiveo & 114 & 5.0 & $272 \pm 1$ & -0.11\\
\ufiveone & 162 & 1.1 & $261 \pm 5$ & -0.12\\
\ueight & 143 & 4.0 & $243 \pm 1$ & -0.03\\
\uten & 45 & 1.1 & $250 \pm 2$ & -0.03\\
\uone & 61 & 4.8 & $280 \pm 9$ & 0.02\\
\enddata
\tablecomments{Col. (2):  Position angle of long axis of the slit in optical spectroscopy. Col. (3): Best-fitting value of $r_{pe}$ (eq.~\ref{fast1:a2}) to folded optical RCs. Col. (4): \Vrot\ measured from hybrid optical+\hi\ RC, corrected for projection effects using $i_d$ from \Iband\ photometry. Col. (5): Logarithmic outer RC slope, measured between $2/3\,r_{235}^o$ from Table~\ref{fast1_tab:iband} and the outermost \hi\ RC point.}
\end{deluxetable}

 
\begin{figure}
\epsscale{1.2}
\plotone{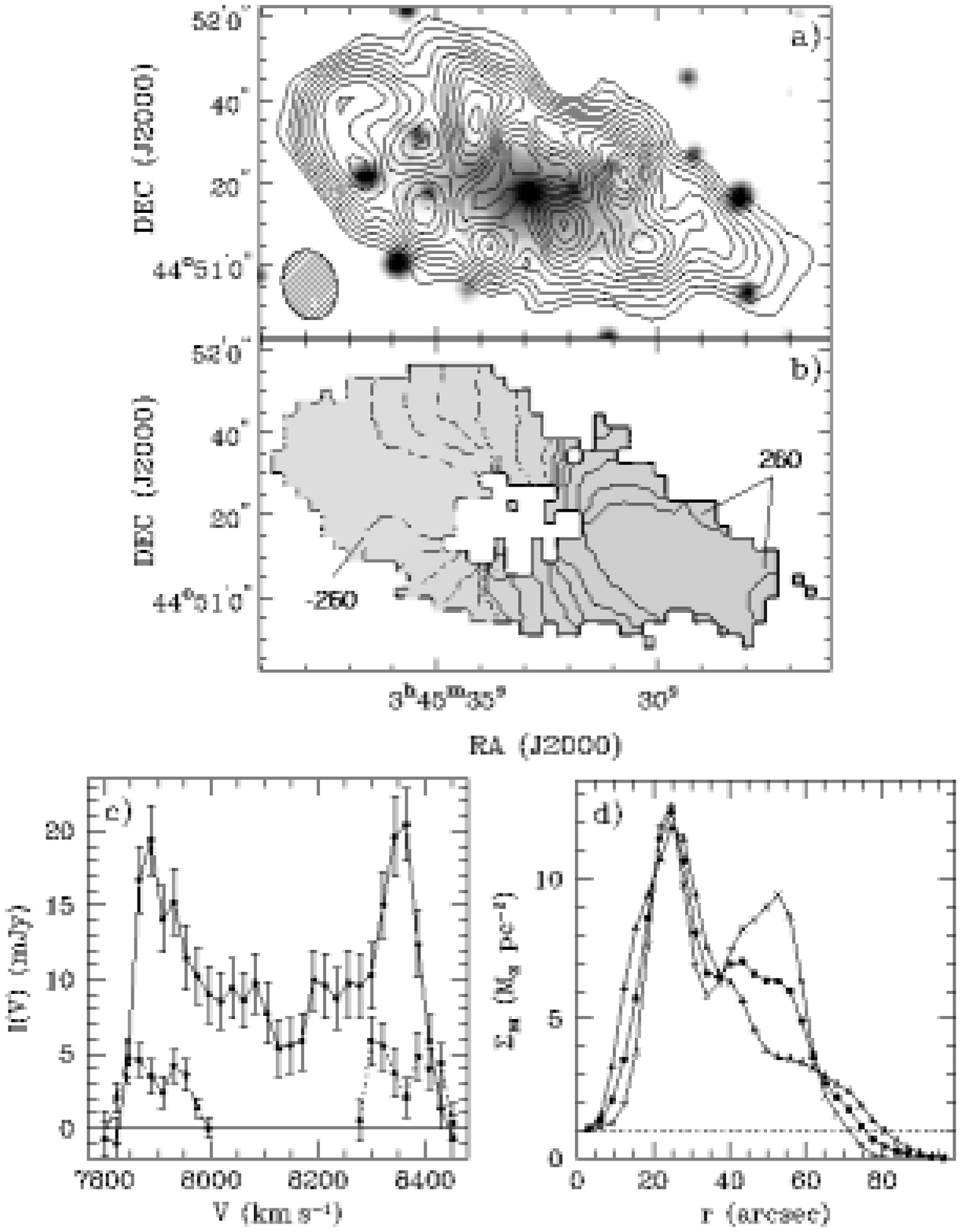}
\caption{Same as Fig.~\ref{fast1_fig:1324summ}, but for \utwo. In $(a)$, \hi\ contours range from $4.7 \times 10^{20}$ to $2.4 \times 10^{21}\,\cm$ in $1.4 \times 10^{20}\,\cm$ intervals, overplotted on a DSS image.}
 \label{fast1_fig:2849summ}
\end{figure}

 
\begin{figure}
\epsscale{1.2}
\plotone{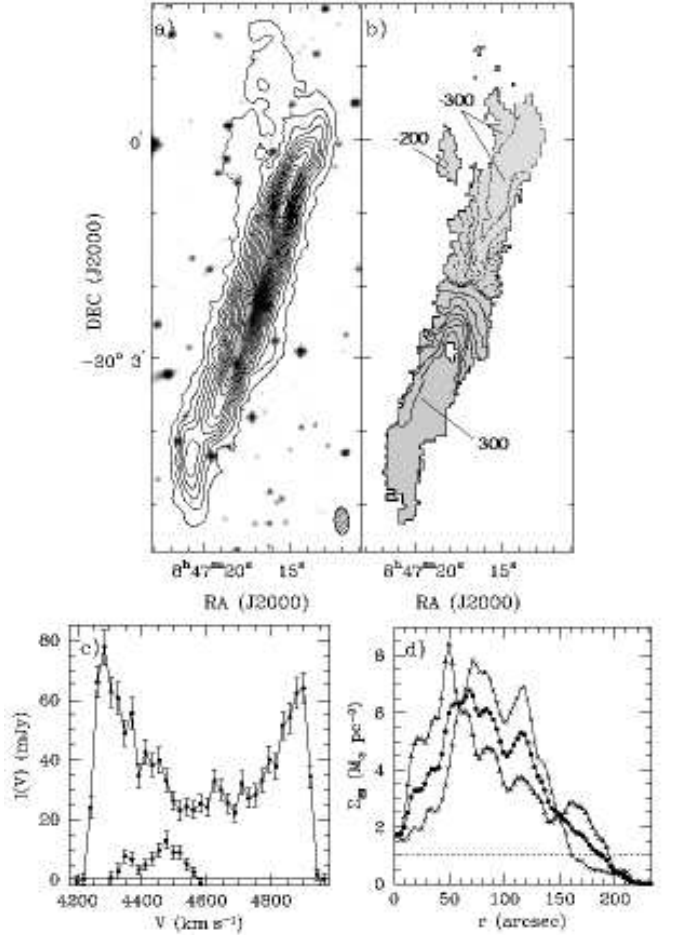}
\caption{Same as Fig.~\ref{fast1_fig:1324summ}, but for ES0~563G21. In $(a)$, naturally weighted \hi\ contours range from $3.4 \times 10^{20}$ to $1.8 \times 10^{22}\,\cm$ in $1.1 \times 10^{21}\,\cm$ intervals, overplotted on a DSS image.}
 \label{fast1_fig:563summ}
\end{figure}
 

\begin{figure}
\epsscale{1.2}
\plotone{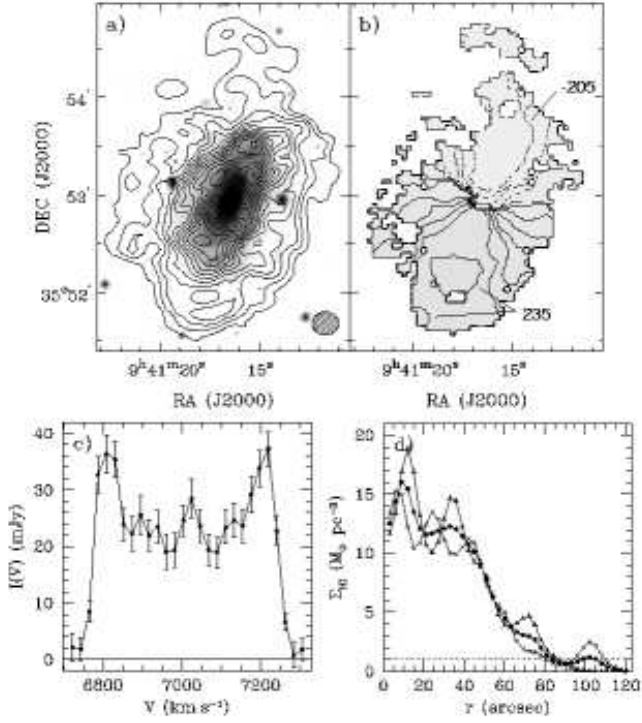}
\caption{Same as Fig.~\ref{fast1_fig:1324summ}, but for \ufiveone. In $(a)$, \hi\ contours range from $1.7 \times 10^{20}$ to $2.8 \times 10^{21}\,\cm$ in $1.7 \times 10^{20}\,\cm$ intervals, overplotted on a DSS image.}
 \label{fast1_fig:5166summ}
\end{figure}
 

\begin{deluxetable*}{ccccccc}
\tablecaption{\hi\ Observing and Map Parameters \label{fast1_tab:HIobs}}
\tablewidth{0pt}
\tablehead{ \colhead{NGC/IC} & \colhead{TOS} & \multicolumn{2}{c}{Natural Weighting} & \multicolumn{2}{c}{Uniform Weighting} & \colhead{Overlap}\\
 & & \colhead{Beam @ $\phi$} & \colhead{$\bar{\sigma}$} & \colhead{Beam @ $\phi$} & \colhead{$\bar{\sigma}$} & \\
 & \colhead{(min)} & \colhead{(\arcsec\ $\times$ \arcsec\ @ $^{\circ}$)} & \colhead{(mJy/beam)} & \colhead{(\arcsec\ $\times$ \arcsec\ @ $^{\circ}$)} & 
\colhead{(mJy/beam)} & \\
\colhead{(1)} & \colhead{(2)} & \colhead{(3)} & \colhead{(4)} & \colhead{(5)} & \colhead{(6)} & \colhead{(7)}
}
\startdata
\none & 264 & 23$\times$17 @ -10 &  0.36 & 17$\times$14 @ -14 & 0.45 & 4\\
\utwo & 273 & 22$\times$17 @ 6 & 0.36 & 17$\times$13 @ 13  & 0.43 & 2\\
\eso & 171 & 22$\times$12 @ 3 & 0.48\tablenotemark{a}  & 13$\times$8 @ 74 & 0.57\tablenotemark{a} & 0\\
\ufiveo & 307 & 18$\times$17 @ -14 &  0.25 & 14$\times$14 @ -23 & 0.30 & 4\\
\ufiveone & 206 & 21$\times$18 @ 82 & 0.33  & 16$\times$14 @ -87 & 0.40 & 8\\
\ueight & 243 & 20$\times$15 @ -7 & 0.31 & 15$\times$13 @ -9 & 0.36 & 6\\
\uten & 378 & 18$\times$17 @ -77 & 0.28 & 14$\times$13 @ -71 & 0.33 & 2\\
\uone & 314 & 20$\times$16 @ -10& 0.28 & 15$\times$13 @ -9 & 0.32 & 3\\ 
\enddata
\tablenotetext{a}{$\sigma$ at $v=4389$\kms\ is 10\% higher than this value; see text.}
\tablecomments{Col. (2): Integration time on-source, before editing. Cols. (3) - (4): Synthesized beam size and position angle, and mean rms map noise per channel in the naturally weighted cubes.  Cols. (5) - (6): Synthesized beam size and position angle, and mean rms map noise per channel in the uniformly weighted maps. Col. (7): Number of overlapping channels near data cube center; see text.}
\end{deluxetable*}

\subsubsection{\hi\ Morphologies and Global Properties}
\label{fast1:global}

 The contours in Figs.~\ref{fast1_fig:1324summ}a--\ref{fast1_fig:11455summ}a show total intensity (zeroth moment) maps for the fast rotators, obtained by summing the detected emission along the frequency axis. To reduce the noise in the maps we use blanked versions of the data cubes shown in Appendix~\ref{fast1:ind}: we first smooth the data to half the spatial resolution listed Table~\ref{fast1_tab:HIobs} and blank the full resolution cube at all locations with emission less than the mean rms $\bar{\sigma}$ of the blanked cubes. We then flag the resulting full-resolution cube interactively, keeping only emission associated with the galaxy of interest and appearing in two consecutive channels. The \hi\ contours in Figs.~\ref{fast1_fig:1324summ}a--\ref{fast1_fig:11455summ}a are overplotted on archived \Iband\ images where available (see \S\ref{fast1:iband}), or else on DSS images. 

 Figs.~\ref{fast1_fig:1324summ}b--\ref{fast1_fig:11455summ}b show velocity fields for each of the fast rotators; their derivation is discussed in detail in \S\ref{fast1:kin}.

 Figs.~\ref{fast1_fig:1324summ}c--\ref{fast1_fig:11455summ}c show the line intensities as a function of heliocentric radial (optical) velocity for the fast rotators and the detected companions. The 1$\sigma$ error bars on each point include contributions from $\bar{\sigma}$ over the emission region as well as a 5\% calibration uncertainty.  The integrated line intensities \ints\ obtained by summing these global profiles along the velocity axis are given in col. (2) of Table~\ref{fast1_tab:properties}. 

 For each system, we compute the raw velocity width $W \sin i$ by fitting straight lines to the datapoints on the outer edges of the global profile horns, and measuring the width from these fits at 50\% of the horn peak. We correct for instrumental broadening by subtracting one channel width from this measurement, in accordance with the recommendation of Springob \etal (2005) from their simulations. Estimates of $W \sin i$ for each fast rotator and the corresponding profile mid-points $V_{sys}$ are in Table~\ref{fast1_tab:properties}. The errors on $W \sin i$ correspond to half the difference between the listed values and those measured at 20\% of the profile peaks using the same technique. We do not correct for interstellar medium (ISM) turbulence, estimated to be on the order of a few kilometers per second and therefore small compared to both $W \sin i $ and its uncertainty (Tully \& Fouqu\'e 1985; Lavezzi \& Dickey 1997; Springob \etal 2005). In general there is good agreement with published single-dish values (see Appendix~\ref{fast1:ind}), and we make no short-spacings corrections to our data. 
 
 The \hi\ mass $M_{HI}$ obtained from the global profile is given in col. (5) of Table~\ref{fast1_tab:properties}, and is computed assuming an optically thin disk:
\begin{equation}
M_{HI} = (2.36 \times 10^5) D^2 \ints \,\,\Msun\,\,\,,
\label{fast1:b}
\end{equation}
where $D$ is in Mpc and \ints\ is in Jy$\,\rm{km\,s^{-1}}$. 

 Radial surface density profiles $\Sigma_{HI}(r)$ are typically obtained by averaging the \hi\ brightness in concentric rings as in Warner \etal (1973), but this method yields erroneous results for systems that are poorly resolved along their major or minor axes (Bosma 1978; Swaters \etal 2002). We therefore adopt the method of Warmels (1988) and integrate the total intensity map for each system along the \Iband\ minor axis to produce ``strip integrals.'' A beam-corrected, face-on estimate of $\Sigma_{HI}(r)$ is then derived from the strip integrals under the assumption of axisymmetry, using Lucy's (1974) iterative deconvolution technique for an initial guess of the distribution shape (see also Swaters \etal 2002).  The resulting $\Sigma_{HI}(r)$ for each system is shown in Figs.~\ref{fast1_fig:1324summ}d -- \ref{fast1_fig:11455summ}d for the approaching side (crosses), the receding side (triangles), and the average of both sides (points) of the \hi\ intensity map in Figs.~\ref{fast1_fig:1324summ}a -- \ref{fast1_fig:11455summ}a. We adopt the outermost point at which $\Sigma_{HI}(r)=1 \,M_{\odot}\,\textrm{pc}^{-2}$ in the averaged profile as the \hi\ radius $r_{HI}$. The ratio $r_{HI}/r_{opt}$ for each system is given in Table~\ref{fast1_tab:properties}.


The total dynamical mass is computed assuming spherical symmetry:
\begin{equation}
M_T = (6.78 \times 10^4)r_{HI} D \,\Vrot^2 \,\Msun\,\,\,,
\label{fast1:c}
\end{equation}
where $r_{HI}$ is in arcminutes and \Vrot\ is in kilometers per second. We note that collisionless \lcdm\ halos are typically oblate with intrinsic axial ratios $0.5 \lesssim q \lesssim 0.7$ (e.g. Jing \& Suto 2002). However, a mild halo ellipticity has little impact on the derivation of $M_T$ (Lequeux 1983), and therefore eq.~\ref{fast1:c} is adequate given the uncertainties in the input parameters.

 An estimate for the dynamical mass is obtained from eq.~\ref{fast1:c} for the approaching and receding sides of each fast rotator separately. We define $r_{HI}$ such that $\Sigma_{HI}(r_{HI})=1 \,M_{\odot}\,\mathrm{pc}^{-2}$, and take \Vrot\ as the plateau of the hybrid RCs near $r_{opt}$ (\S\ref{fast1:kinboth}). The results are then averaged to produce $M_T$, listed in Table~\ref{fast1_tab:properties}. We adopt an uncertainty on $M_T$ equal to the sum in quadrature of half the difference between $M_T$ for the approaching and receding sides, and the uncertainty obtained by propagating errors in $r_{HI}$ and \Vrot\ through eq.~\ref{fast1:c}.  The average value of \Vrot\ for the sample galaxies is given in Table~\ref{fast1_tab:kin}, where error bars represent half the difference between values obtained for the approaching and receding sides.

 We also list the parameters $M_{HI}/M_{T}$, $\sigma_{HI}=M_{HI}/\pi r_{HI}^2$ and $\sigma_{T}=M_T/\pi r_{HI}^2$ in Table~\ref{fast1_tab:properties}. They demonstrate that despite their large absolute scale, the fast rotators have the same average properties as normal spiral galaxies: while $M_{HI}$ and $M_{T}$ for all of the fast rotators exceed the 75$^{\textrm{th}}$ percentiles of the volume-limited RC3-LSc sample of Roberts \& Haynes (1994), the values of $M_{HI}/M_{T}$, $\sigma_{HI}$ and $\sigma_{T}$ lie within the expected ranges for their morphological types\footnote{We adjust the values in Table~\ref{fast1_tab:properties} to account for the different cosmology and parameter definitions adopted in  Roberts \& Haynes (1994) before making the comparison.}. 

 The global properties of the detected companions are given in Table~\ref{fast1_tab:propertiesc}. 
Estimates of \ints, $V_{sys}$, \wsini, and $M_{HI}$ are computed in the same manner as for the fast rotators, adopting $D$ of the parent system in eq.~\ref{fast1:b}. For $M_T$ we approximate $r_{HI} = 1.5 \times a/2$ and \Vrot$= W \sin i/\cos^{-1}(b/a)$, where $a$ and $b$ are the optical major and minor axis diameters. Given the uncertain morphologies, poor resolution and limited S/N of the detected companions, $M_T$ should be regarded as only a rough approximation to the dynamical masses of these systems.

\subsubsection{\hi\  Kinematics}
\label{fast1:kin}

 Rather than adopting a traditional intensity weighting (or first moment; Warner \etal 1973) scheme to derive the velocity fields in Figs~\ref{fast1_fig:1324summ}b--\ref{fast1_fig:11455summ}b, we use G04's modified envelope tracing (MET) method.  We describe the technique, its advantages over moment analyses, and its application to our data here. 
 For clarity in what follows, let $V_{mom1}(\alpha,\delta)$ and $V_{MET}(\alpha,\delta)$ refer to radial velocities at a sky position $(\alpha,\delta)$ derived using the first moment and MET techniques, respectively. Quantities as a function of $r$ instead of $(\alpha,\delta)$ are measured along the major axis of the system, and the kinematic center is at $r=0$. We use $V_{obs}(r)$ to denote the radial velocity adopted at $r$. For an axisymmetric, rotationally supported disk, $V_{obs}(r)=V_c(r)\sin{i}$, where $V_c(r)$ is the circular velocity of the system at $r$. We wish to derive $V_{obs}(r)$ values that approach $V_c(r)\sin{i}$.

 First moment velocities are obtained by weighting each contribution by its intensity $I(\alpha,\delta,v)$:
\begin{equation}
V_{mom1}(\alpha,\delta)=\frac{\int I(\alpha,\delta,v) v \mathrm{d} v}{\int I(\alpha,\delta,v) \mathrm{d} v}\,\,\,,
\label{fast1:d}
\end{equation}
 where $v$ represents the frequency or velocity axis. When the line profiles $I(\alpha,\delta,v)$ are symmetric about the profile peaks the first moment yields reliable estimates of $V_c(r)\sin{i}$, as do methods that fit gaussians or find centroids (see Sofue \& Rubin 2001 for a review). In highly inclined or poorly resolved disks, however, a single line of sight will intercept lower-velocity emission in addition to gas at $V_c(r)\sin{i}$: this results in profiles that are skewed toward $V_{sys}$, biasing $V_{mom1}(\alpha,\delta)$ lower than $V_c(\alpha,\delta)$. 

\begin{figure}
\epsscale{1.2}
\plotone{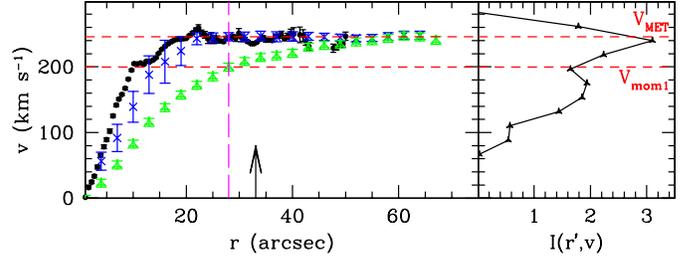} 
\caption{Left: comparison between $V_{mom1}(r_i)$ (triangles) and $V_{MET}(r_i)$ (crosses) for the receding side of \ueight. The filled circles show the velocities derived from long-slit spectroscopy, for which beam-smearing effects are negligible. 
The vertical long-dashed line shows the location $r^{\prime}$ at which $I(r^{\prime},v)$ in the right panel is extracted (units are arbitrary), and the horizontal dashed lines show $V_{mom1}(r^{\prime})$ (lower dashed line) and $V_{MET}(r^{\prime})$ (upper dashed line). The solid arrow denotes \ropt. The MET technique mitigates the impact of beam smearing in the \hi\ kinematics, and there is good agreement between $V_{MET}(r_i)$ and optical tracers near \ropt. See the electronic edition of the Journal for a color version of this figure.}
 \label{fast1_fig:met}
\end{figure}

 The impact of this bias in our data is illustrated in Fig.~\ref{fast1_fig:met}. In the left part of the figure, the points show different estimates of the raw RC for the receding side of \ueight: the triangles denote $V_{mom1}(r_i)$ computed from eq.~\ref{fast1:d}. The solid circles show $V_{obs}(r_i)$ derived from the optical spectroscopy described in \S\ref{fast1:halpha}, for which beam smearing effects are negligible. In the absence of observing and data reduction biases, we expect $V_{obs}(r_i)$ derived from \hi\ to resemble that derived from \ha\ in the region probed by both tracers. The right part of the figure shows the spectral profile $I(r^{\prime},v)$ extracted at the location corresponding to the long-dashed vertical line in the left panel. The lower horizontal dashed line denotes $V_{mom1}(r^{\prime})$ derived for that profile. This value is clearly biased towards $V_{sys}$ (= 0 in this plot) relative to the profile peak, and as a result there is a significant discrepancy between $V_{mom1}(r^{\prime})$ and $V_{obs}(r^{\prime})$ estimated from optical spectroscopy. 

 We therefore adopt the MET method of G04 to derive velocity fields for the fast rotator sample.
For emission tracing the kinematics of an inclined rotating disk, the extreme-velocity side of $I(\alpha,\delta,v)$ (that farthest from $V_{sys}$) represents $V_c(\alpha,\delta)$ modulated by the telescope response and ISM turbulence. An estimate of the latter can therefore be derived from a ``terminal velocity'' $V_t(\alpha,\delta)$, which G04 define as the velocity at half-maximum of a half Gaussian fit to the extreme-velocity side of the profile. The velocity field is then given by
\begin{equation}
V_{MET}(\alpha,\delta)=V_t(\alpha,\delta) - 0.5\sqrt{ (\Delta V_{ISM})^2 + (\Delta V_{obs})^2 }\,\,\,. 
\label{fast1:e}
\end{equation}
 In eq.~\ref{fast1:e}, $\Delta V_{ISM}$ accounts for turbulent broadening of the ISM and $\Delta V_{obs}$ corrects for the instrumental broadening of the system. Following G04 we assume a constant velocity dispersion $\sigma_{ISM}=10$\kms\ with $\Delta V_{ISM} = 2.35\sigma_{ISM}$ and set $\Delta V_{obs}$ to the spectral channel width in kilometers per second. For our data sets, $\Delta V_{ISM}$ and $\Delta V_{obs}$ have similar amplitudes, and the $V_{MET}(\alpha,\delta)$ values obtained are insensitive to changes in $\sigma_{ISM}$ in the range $7\,\rm{km\,s^{-1}} \lesssim \sigma_{ISM} \lesssim 12\,\rm{km\,s^{-1}}$ (Kamphuis 1993).

 Phenomena other than turbulence and spectral resolution can alter the extreme-velocity side of the profiles. Some of these stem from the galaxy itself, such as outflows or high velocity cloud (HVC) analogs. These ``internal'' factors are of little concern in our sample, since we are interested mainly in the outer \hi\ disk and do not have the sensitivity to detect putative HVCs (Pisano \etal 2004). One should also account for spatial beam smearing in regions in which the RC derivative is large: this will broaden the extreme-velocity side of the profile, since along a line of sight gas at larger $r$ [higher $V_c(r)$] will be included via convolution by the synthesized beam. The $V_{MET}(r)$ derived from eq.~\ref{fast1:e} then is, in principle, an upper limit to $V_c(r)\sin{i}$ (e.g. Sancisi \& Allen 1979). However, minor axis resolution effects are so severe in the inner regions of the fast rotator disks that eq.~\ref{fast1:e} {\it under-estimates} $V_c(r)\sin{i}$ even when beam smearing is neglected. This problem was also noted by G04, who adjust the inner points of their RCs by hand. Since the RC gradients in the outer disks are small for the fast rotators (Figs.~\ref{fast1_fig:edgeons}--\ref{fast1_fig:intinc}), and since beam-smearing corrections are necessarily ad-hoc (e.g. Braun 1997; G04) and will further depress $V_{MET}(\alpha,\delta)$ in the inner regions, we apply no such corrections to the velocity fields. We note that \ion{H}{1} points within the inner 1.5 beams of the galaxy center are ultimately dropped in the hybrid RC derivation (see \S\ref{fast1:hybrid}); our neglect of this effect thus has no impact on the final results.

 The application of the MET method to our data is demonstrated in Fig.~\ref{fast1_fig:met}. The crosses show $V_{MET}(r_i)$ along the peak of the intensity distribution, and the upper horizontal dashed line shows the value computed from the line profile in the right panel. 
At large $r$ there is little low-velocity gas along the line of sight, and  $V_{MET}(r_i) \simeq V_{mom1}(r_i)$. Within \ropt, it is clear that $V_{MET}(r_i)$ resembles the $V_{obs}(r_i)$ from optical spectroscopy (filled circles) much more closely than does $V_{mom1}(r_i)$ (triangles). 
In particular, there is good agreement between $V_{MET}(r)$ and $V_{obs}(r)$ measured in the optical near $r_{opt}$, leading to a smooth transition between the kinematics derived from the two tracers. 

\begin{deluxetable*}{cccccccccc}
\tablecaption{\hi\ Properties of the Detected Companions \label{fast1_tab:propertiesc}}
\tabletypesize{\scriptsize}
\tablewidth{0pt}
\tablehead{\colhead{Name} & \colhead{Parent} & \multicolumn{2}{c}{Separation} & \colhead{$a \times b$} & \colhead{\ints} & \colhead{$W \sin i$} & \colhead{$V_{sys}$} & \colhead{$M_{HI}$} & \colhead{$M_T$}\\
 &  & \multicolumn{2}{c}{(\arcmin)} & {(\arcmin\ $\times$ \arcmin)}  & \colhead{(Jy$\,$\kms)} & \colhead{(\kms)} & \colhead{(\kms)} & \colhead{($\times 10^9\,$\Msun)} &  \colhead{($\times 10^9\,$\Msun)}\\
\colhead{(1)} & \colhead{(2)} & \multicolumn{2}{c}{(3)} & \colhead{(4)} & \colhead{(5)} & \colhead{(6)} & \colhead{(7)} & \colhead{(8)} & \colhead{(9)}
}
\startdata
HI 034538+444639 & \utwo & 4.7 & (S) & $0.15 \times 0.05$ & $0.56 \pm 0.07$ & $119 \pm 16$ & 7897 & $1.7 \pm 0.2$ & $3.8 \pm 0.7$\\
HI 034553+445126 & \utwo & 3.5 & (E) & $0.11 \times 0.06$ & $0.7 \pm 0.1$ & $129 \pm 7$ & 8365 & $2.2 \pm 0.3$ & $3.3 \pm 0.3$ \\
ESO 563G22 & \eso &  5.9 & (NE) & $1.1 \times 0.2$ & $1.7 \pm 0.2$ & $173 \pm 21$ & 4429 & $2.0 \pm 0.2$ & $30 \pm 5$\\
KUG~1634+392 & \uten &   5.7 & (N) & $0.3 \times 0.3$ & $0.29 \pm 0.05$ & $42 \pm 19$ & 8975 & $1.2 \pm 0.2$ & $\ge 0.9 \pm 0.6$\tablenotemark{a}\\
HI 163617+390413 & \uten & 4.0 & (NW) & $0.13\times0.08$ & $0.83 \pm 0.07$ & $99 \pm 11$ & 9085 & $3.3 \pm 0.3$ & $3.13 \pm 0.5$ \\
CGCG~341-027 & \uone & 3.4 & (NW) & $0.55 \times 0.2$ & $0.78 \pm 0.08$ & $263 \pm 18$ & $5146$ & $1.1 \pm 0.1$ & $42 \pm 4$ \\ 
\enddata
\tablenotetext{a}{No correction for inclination applied since $a=b$; lower limit.}
\tablecomments{Col. (1): Source name. Optical identifications are used where possible. For previously uncatalogued sources, names assigned from J2000 coordinates of the \hi\ centroid in total intensity maps. Col. (2): Parent galaxy. Col (3): Projected separation from parent galaxy. The direction of the companion relative to the parent is given in parentheses. Col. (4): Optical dimensions from literature if available, otherwise estimated from \Iband\  (HI 163617+390413) or DSS (HI 034538+444639 and HI 034553+445126) images. Col. (5): Integrated line intensity. Col. (6): Velocity width at 50\% of the peak, corrected for instrumental broadening.  Col. (7): Integrated profile mid-point. Col.(8): \hi\ mass, computed from eq.~\ref{fast1:b} using $D$ for parent galaxy. Col. (9): Total dynamical mass, computed from eq.~\ref{fast1:c} (see text). }
\end{deluxetable*}

 The MET velocity fields derived within the lowest contour in Figs~\ref{fast1_fig:1324summ}a--\ref{fast1_fig:11455summ}a are shown in Figs~\ref{fast1_fig:1324summ}b--\ref{fast1_fig:11455summ}b (shaded regions). The isovelocity contours are spaced at $40\,$\kms\ intervals: dotted lines denote the approaching side relative to $V_{sys}$ in Table~\ref{fast1_tab:properties}, and solid lines denote the receding side. Pixels within the \hi\ disk with no $V_{MET}$ estimates correspond to regions in which a reliable Gaussian fit to the extreme-velocity side of the profile could not be obtained. This is particularly  prevalent near the galaxy centers, where beam smearing effects are large and the \hi\ surface density is low (Figs~\ref{fast1_fig:1324summ}d--\ref{fast1_fig:11455summ}d). 

 We note that while the premise of the MET technique is physically motivated, the details of its application are somewhat arbitrary. In particular, the fraction of the fitted Gaussian peak adopted as $V_t(\alpha,\delta)$ and the form of the corrections applied are empirically derived (G04). In light of these uncertainties, \citet{kregel04b} argue that iterative least-squares modeling of position-velocity (PV) diagrams yields superior disk kinematics to the method adopted here. However, the resolution and dynamic range in our data are insufficient to assure convergence of the technique, and it can only be applied along the major axes of edge-on galaxies; as such, we do not consider it here. Moreover, the excellent agreement between the optical and \hi\ RCs at $r_{opt}$, as well as that between the observed and simulated \hi\ morphologies discussed in \S\ref{fast1:kinboth}, suggest that the MET technique is indeed recovering $V_c(r) \sin{i}$, and that further refinements to eq.~\ref{fast1:e} are unnecessary.

\section{RC derivation}
\label{fast1:kinboth}

  In this section we derive \hi\ RCs from the velocity fields in  Figs.~\ref{fast1_fig:1324summ}b--\ref{fast1_fig:11455summ}b and combine them with the optical RCs of \S\ref{fast1:halpha}. We note that slightly different methods are adopted to obtain \hi\ RCs of the highly inclined and intermediate-$i$ systems in the sample, and we discuss each case in turn below. Since $i_d=78^{\circ}$ for \none, its kinematics can be determined using either approach: we verify that both produce the same result. Because $r_{pe}$ for \none\ is similar to those of \utwo, \ufiveone\ and \uten\ (Table~\ref{fast1_tab:kin}), however, we group it with the intermediate-$i$ systems below. 

\subsection{\hi\ RCs for the Highly Inclined Systems}
\label{fast1:highi}

 We determine $V_{obs}(r)$ for the highly inclined systems \eso, \ufiveo, \ueight\ and \uone\  by averaging $V_{MET}(\alpha,\delta)$ from Figs.~\ref{fast1_fig:1324summ}b--\ref{fast1_fig:11455summ}b within a synthesized beam of the intensity distribution peak in Figs.~\ref{fast1_fig:1324summ}a--\ref{fast1_fig:11455summ}a. This strategy is similar to the warped-MET technique described by G04 and naturally follows the disk kinematics along the ridge of any warp that may be present in the disk. For \eso, \ufiveo, and \uone\ we estimate the warp parameters by tracing $\theta$ of the intensity distribution peak, as in \citet{gar02}. We also adopt their definition of the warp angle $\alpha$ (their eq.~4) as that between the major axis and a line connecting the galaxy center and the outermost measured point in the warp. The derived $\alpha$ values are accurate to $\sim 5^{\circ}$, while the warp radii $r_w$ are accurate to a few kiloparsec. The warp parameters for these systems are given in Table~\ref{fast1_tab:properties} (see also \S\ref{fast1:ind}), and $r_w$ is shown in Fig.~\ref{fast1_fig:edgeons}.  

 To verify that the RCs, $\alpha$ and $r_w$ derived are consistent with the \hi\ morphologies of each system, we simulate observations of \hi\ disks with the derived properties and compare them to the channel maps in Figs.~\ref{fast1:563chans},~\ref{fast1:5010chans},~\ref{fast1:8220chans} and \ref{fast1:11455chans}. Each galaxy is modeled by concentric rings with a user-supplied $i$, $\theta$, $V_c$ and $\Sigma_{HI}$. All rings have systemic velocities $V_o=V_{sys}$ from Table~\ref{fast1_tab:properties} and centers $(\alpha_o,\delta_o)=(\alpha_c,\delta_c)$ from Table~\ref{fast1_tab:basic}. We model the approaching and receding sides of each fast rotator separately, setting $V_c(r_i)=V_{obs}(r_i)/\sin{i}$ and using $\Sigma_{HI}(r_i)$ from Figs.~\ref{fast1_fig:1324summ}d--\ref{fast1_fig:11455summ}d at the center $r_i$ of each ring. We set $\theta=\theta_d$ for $r_i<r_w$, and for $r_i>r_w$ we introduce a smooth variation in $\theta$ consistent with $\alpha$. We set $i=i_d$ at all $r_i$ and adopt $\sigma_{ISM}=10\,\kms$, as well as an exponential \hi\ scale-height of $1\,$kpc. 

 We find good agreement between the simulated data cubes and the channel maps in Figs.~\ref{fast1:1324chans}--\ref{fast1:11455chans} for all of the highly inclined systems. In particular, setting $i=i_d$ reproduces the observed \hi\ morphologies even beyond $r_w$: $i$ of the detected warps thus approaches that in the corresponding disk.  We experiment with different input $i$ beyond $r_w$ and find that setting $i \lesssim 80^{\circ}$ in the simulations fails to recover the \hi\ channel maps in the warp regions. There is therefore no ambiguity between $V_{obs}(r)$ and $i$ in these systems. 
Our simulations also demonstrate that the derived \hi\ surface densities, kinematics, and disk geometries for the highly inclined fast rotators self-consistently recover the morphology of the detected emission.  

\subsection{\hi\ RCs for the Intermediate-$i$ Systems}
\label{fast1:inti}

 When \hi\ disks at intermediate $i$ are sufficiently resolved, RCs can be derived by fitting a series of concentric rings to the observed velocity field and by solving for $V_c$, $V_o$, $i$, $\theta$ and $(\alpha_o,\delta_o)$ in each ring (Begeman 1989). We perform a tilted-ring analysis of the intermediate-i systems,  first allowing all six parameters to vary and then fixing $V_o$ and $(\alpha_o,\delta_o)$ and re-fitting for $V_c$, $i$ and $\theta$. The values of $V_o$, $(\alpha_o,\delta_o)$ and $\theta$ obtained from the fits are comparable to $V_{sys}$ (Table~\ref{fast1_tab:properties}), $(\alpha_c,\delta_c)$ (Table~\ref{fast1_tab:basic}) and $\theta_d$ (Table~\ref{fast1_tab:iband}), respectively. However, the best-fitting $i$ values are lower than $i_d$ (Table~\ref{fast1_tab:iband}) by an average of 15\%, and $V_c(r)$ is $\sim 10\%$ larger than that from the optical kinematics and from major-axis slices. In addition, simulated data cubes of systems with these properties produce models with spatial and spectral extents that are too large when compared to observations. These discrepancies arise from the low angular resolution of the latter along the minor axes of the intermediate-$i$ systems (Figs.~\ref{fast1_fig:1324summ}a--\ref{fast1_fig:11455summ}a): beam smearing blends major axis gas with lower velocity off-axis material in this direction, and broadens the spatial extent of the emission. This leads to an underestimate of $i$ and a corresponding overestimate of $V_c$ in the tilted-ring models. The biases on the ring properties are analogous to those encountered when ellipse fits are used to determine $\Sigma_{HI}(r)$ (see \S\ref{fast1:HI}).

 We therefore derive the RCs for \none, \utwo, \ufiveone\ and \uten\ in a manner similar to that for the high $i$ systems, averaging $V_{MET}(\alpha, \delta)$ from Figs.~\ref{fast1_fig:1324summ}b--\ref{fast1_fig:11455summ}b in a slice the width of a synthesized beam oriented at $\theta_d$ (Table~\ref{fast1_tab:iband}). We thus implicitly assume that the major axis of the \hi\ disk lies within a beam of the \Iband\ major axis. These assumptions do appear to be justified given the \hi\ morphologies and kinematics in Figs~\ref{fast1_fig:1324summ}--\ref{fast1_fig:11455summ}. 

  To verify the plausibility of the $V_{obs}(r)$ obtained, we simulate observations of each intermediate $i$ fast rotator in a manner similar to that for the highly inclined systems. We model the approaching and receding sides of each galaxy separately, adopting  $\Sigma_{HI}(r)$ from Figs.~\ref{fast1_fig:1324summ}d--\ref{fast1_fig:11455summ}d and $\theta=\theta_d$ at all $r$. Unlike the near-edge-on case, however, at intermediate $i$ there is an ambiguity between $i$ and the RC amplitude in the outer \hi\ layer. In our sample, $V_{obs}(r)$ for \none, \utwo\ and \ufiveo\ declines beyond \ropt\ on at least one side (see \S\ref{fast1:hybrid}). Moreover, changes in $i$ of only a few degrees have a significant impact on the resulting $V_c(r) = V_{obs}(r)/\sin{i}$: these decreases may thus represent a change in the gravitational potential in these regions or may result from variations in the disk geometry. Note that minor-axis distortions indicative of a warped disk are difficult to detect in our data, due to limited spatial resolution along this axis: we therefore make no attempt to measure warp parameters in the intermediate $i$ systems.

 To investigate the extent to which the \hi\ channel maps can distinguish
between a falling outer RC and a change in the HI disk geometry, we construct two models for each of the galaxies with declining $V_{obs}(r)$. In the first model we set $i=i_d$ at all $r$, and in the second we introduce a smooth decrease (increase for the approaching side of \utwo) in $i$ beyond \ropt\ to produce a flat RC. Details of this second model are given in the notes on individual systems in \S\ref{fast1:ind}, but $0^{\circ} < i_d - i(r_{HI}) < 15^{\circ}$ for all systems.  In general, there is good agreement between the simulated observations and the measured channel maps for {\it both} models (see \S\ref{fast1:ind} for details); the available data thus cannot distinguish between a falling RC due to a change in the gravitational potential beyond \ropt\ and one caused by a variation in $i$ associated with a warp. We discuss this further in \S\ref{fast1:sample}.

\subsection{Hybrid RC Derivation}
\label{fast1:hybrid}

\begin{figure}
\epsscale{1.2}
\plotone{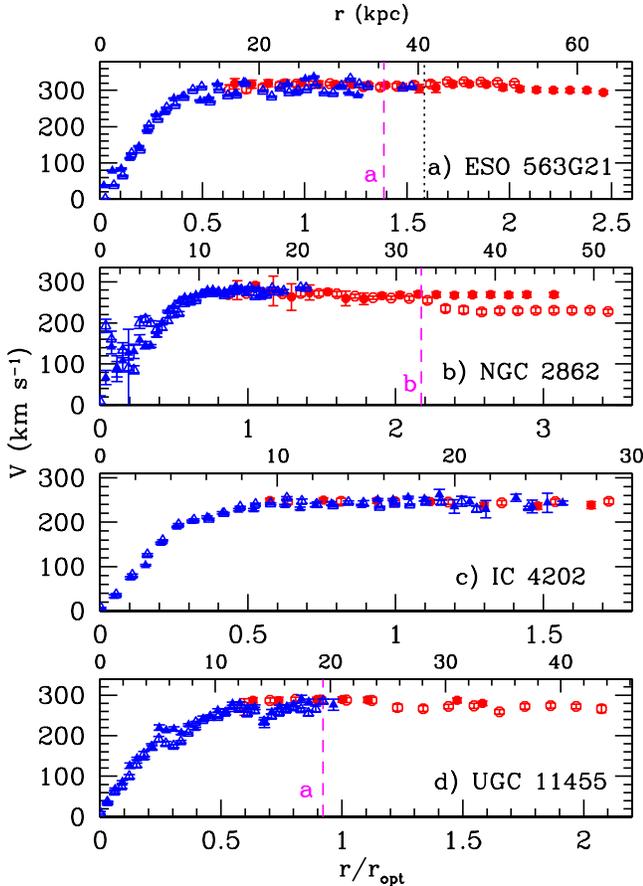}  
\caption{Hybrid \hi/optical RCs [$V_{obs}(r_i)/\sin{i_d}$] for the highly inclined galaxies. In each panel, $r_i$ is in units of $r_{opt}$ on the lower horizontal axis, and in kiloparsecs on the upper horizontal axis. RC points derived from long-slit spectroscopy are denoted by triangles, and those from the \hi\ observations are denoted by circles. Open symbols correspond to the approaching side and filled symbols to the receding side. The dashed vertical lines indicate the measured warp radii, and the label to the lower left of each line distinguishes warps on the (a) approaching or on (b) both sides of the disk. The galaxy name is given in the bottom right corner. For \eso\ in $(a)$, the dotted vertical line shows the approximate starting location of the southeastern feature; see \S\ref{fast1:563notes}. See the electronic edition of the Journal for a color version of this figure.}
\label{fast1_fig:edgeons}
\end{figure}
 
 
\begin{figure}
\epsscale{1.2}
\plotone{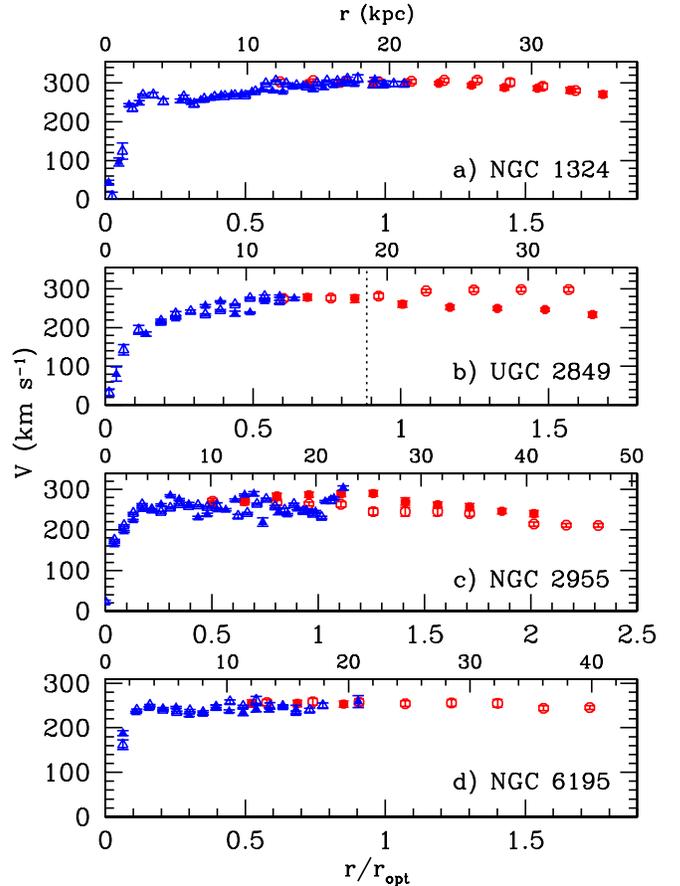} 
\caption{Hybrid \hi/optical RCs [$V_{obs}(r_i)/\sin{i_d}$] for the intermediate-$i$ galaxies; plot details are the same as in Fig.~\ref{fast1_fig:edgeons}. For \utwo\ in $(b)$, the dotted vertical line shows the approximate starting location of the \hi\ emission on the receding side that may be associated with interactions between this galaxy and the detected companions; see \S\ref{fast1:2849notes}. See the electronic edition of the Journal for a color version of this figure.}
\label{fast1_fig:intinc}
\end{figure}

 Further processing of the \hi\ RCs as well as their combination with the optical RCs discussed in \S\ref{fast1:halpha} is done homogeneously for all of the sample galaxies. We fold each \hi\ RC about the point ($r_{min},V_{min}$) that minimizes the difference between the approaching and receding sides within $r_{opt}$. These points are assigned to $r=0, V=0$ in the RC, and they differ by no more than 15\kms\ and 5\arcsec\ from $V_{sys}$ measured in \S\ref{fast1:global} (Table~\ref{fast1_tab:properties}) and the optical center $(\alpha_c,\delta_c)$ (Table~\ref{fast1_tab:basic}), respectively. We assign half the difference between $V_{MET}(r)$ obtained when $V_t(\alpha,\delta)$ in eq.~\ref{fast1:e} is measured at 50\% of the $I(\alpha,\delta,v)$ peak and when it is measured at 20\% of the peak as the error on each \hi\ RC point. The \hi\ RCs are then resampled to 6\arcsec\ (6.2\arcsec\ for \eso), yielding slightly less than two points per synthesized beam.

 We then combine the optical and \hi\ RCs, dropping \hi\ points within 1.5 synthesized beam widths of $r=0$ to avoid beam smearing biases on $V_{obs}(r)$ (\S\ref{fast1:kin}) and correcting $r_i$ of the remaining \hi\ points for beam smearing. We also clip optical RC points that are discrepant by over 50\kms\ and are thus likely the result of non-circular motions or patchy emission regions in the disk. Finally, we correct for galaxy geometry using $i_d$ from Table~\ref{fast1_tab:iband}. This has a negligible effect on the kinematics of the highly inclined systems, but the RC amplitudes and shapes in the intermediate $i$ subsample depend on this choice. 

 The hybrid \hi/optical RCs [$V_{obs}(r_i)/\sin{i_d}$] for the highly inclined galaxies in the fast rotator sample are shown in Fig.~\ref{fast1_fig:edgeons}, and those for the intermediate $i$ systems are in Fig.~\ref{fast1_fig:intinc}. In each panel, the triangles denote points derived from long-slit spectra while the circles denote points obtained from \hi\ observations. Open and filled symbols show the approaching and receding sides, respectively.  In Fig.~\ref{fast1_fig:edgeons} dashed lines indicate $r_w$ in the highly inclined systems with warps, and those on the approaching side (a) or on both sides (b) of the disks are labeled to the lower left of the line.  For \eso\ in Fig.~\ref{fast1_fig:edgeons}a, the dotted line shows the approximate starting location of the southeastern feature; see \S\ref{fast1:563notes} for details.  Beyond the vertical dotted line in Fig.~\ref{fast1_fig:intinc}b, the \hi\ channel maps on the receding side of \utwo\ are not well modeled by either a co-planar or warped disk; it is possible that interactions with its detected companions are distorting the \hi\ layer in this region (see \S\ref{fast1:2849notes}).

 For all of the sample galaxies, there is a smooth transition between the optical and \hi\ RCs as well as excellent agreement between points in the overlap regions near \ropt. This suggests that the ionized and atomic components of the ISM have similar geometries at this location in the underlying galaxy potential.

\begin{figure*}
\figurenum{13}
\epsscale{1.0}
\plotone{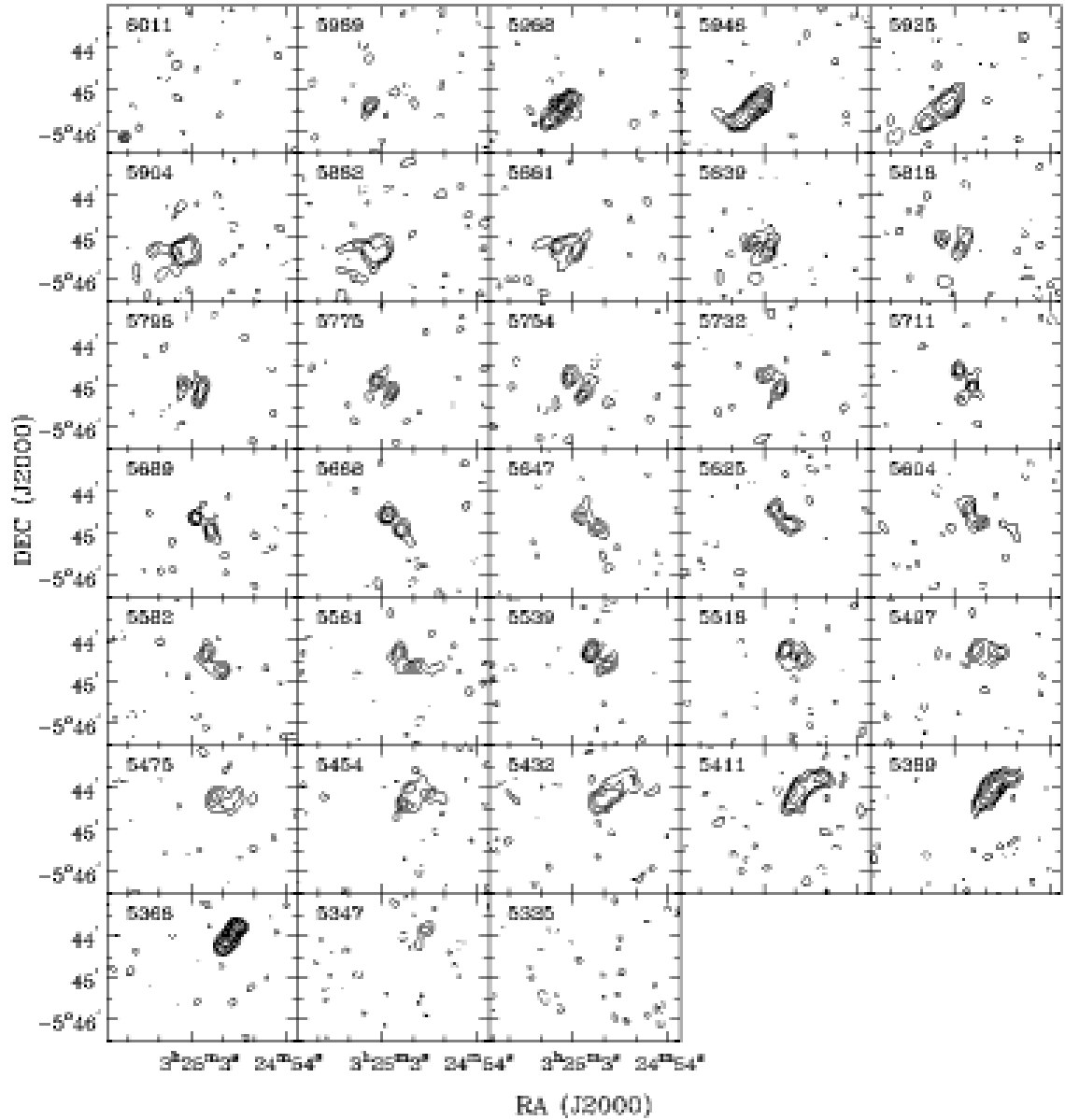}
\caption{Channel Maps. These sample panels show uniformly weighted channel maps for \none. Contours are at $0.45\,\times$ (-3.5, -2, 2 ($2\sigma$), 3.5, 5, 8, 11, 14, 16) mJy$\,$beam$^-1$. In each panel, the heliocentric (optical) radial velocity is in the upper left corner. The synthesized beam is in the lower left corner of the first panel. See the electronic edition of the Journal for Figs.~\ref{fast1:2849chans}--\ref{fast1:11455chans}. }
\label{fast1:1324chans}
\end{figure*}

\begin{figure}
\figurenum{13.2}
\label{fast1:2849chans}
\end{figure}

\begin{figure}
\figurenum{13.3}
\label{fast1:563chans}
\end{figure}

\begin{figure}
\figurenum{13.4}
\label{fast1:5010chans}
\end{figure}

\begin{figure}
\figurenum{13.5}
\label{fast1:5166chans}
\end{figure}

\begin{figure}
\figurenum{13.6}
\label{fast1:8220chans}
\end{figure}

\begin{figure}
\figurenum{13.7}
\label{fast1:10469chans}
\end{figure} 

\begin{figure}
\figurenum{13.8}
\label{fast1:11455chans}
\end{figure}

\section{Discussion and Summary}
\label{fast1:sample}

 We have presented \Iband\ photometry, long-slit optical spectra, and new \hi\ aperture synthesis observations for eight rapidly rotating nearby spiral galaxies. Despite their large scales, the sample galaxies lie on the size, velocity, absolute magnitude and surface brightness scaling relations in the SFI++ database, and their global gas fractions and surface densities are within the ranges expected for their morphological types (\S\ref{fast1:global}; Table~\ref{fast1_tab:properties}). These ``fast rotators'' thus represent the high-mass extreme of the normal, late-type spiral galaxy population. We derive hybrid optical/\hi\ rotation curves (RCs) for each system, and find a good correspondence between the \hi\ rotation velocities derived using G04's Modified Envelope Tracing method and those from optical spectra in regions probed by both tracers (Figs.~\ref{fast1_fig:edgeons}--\ref{fast1_fig:intinc}). In a forthcoming paper we will mass model the kinematics presented here, and estimate corresponding angular momentum distributions as well as spin parameters $\lambda$ for a range of baryon mass-to-light ratios (K. Spekkens \& R. Giovanelli 2006, in preparation).  
 
 In general, the detected \hi\ emission in each galaxy has the morphology of a thin rotating disk with the same geometry as its \Iband\ stellar distribution. However, we find that warps are prevalent beyond \ropt\ in massive spiral galaxies: all of the highly inclined, extended \hi\ disks that we detect are warped on one or both sides, and the channel maps for the intermediate-$i$ systems are also consistent with warped outer layers. Some of these warps  (e.g. \eso\ and \uone) may be excited by companions, which we detect in the vicinity of half the sample galaxies (Table~\ref{fast1_tab:propertiesc}). Interactions with these smaller systems may have also produced the \hi\ features in \utwo\ (Fig.~\ref{fast1_fig:intinc}; \S\ref{fast1:2849notes}) and \eso\ (Fig.~\ref{fast1_fig:edgeons}; \S\ref{fast1:563notes}) that we are unable to recover in our disk models of the observed \hi\ emission. As a counter-example, however, we point out the spectacular symmetric warp in the {\it isolated} system \ufiveo\ (Figs.~\ref{fast1_fig:5010summ},~\ref{fast1:5010chans}): its warp angles are comparable to the largest detected by \citet{gar02} in their sample of 26 edge-on spiral galaxies and are $30\%$ larger than those in their isolated systems. 

 Despite the presence of warps and other features in the \hi\ morphologies of the fast rotators, their kinematics are very regular. \eso\  is an extreme example: we detect prominent \hi\ extensions on either end of the major axis (Fig.~\ref{fast1_fig:563summ}) but find well-ordered rotation throughout the \hi\ layer irrespective of the inclusion of the discrepant emission (\S\ref{fast1:563notes}; Fig.~\ref{fast1_fig:edgeons}a).  There is also excellent agreement between the kinematics of the approaching and receding sides in most systems (exceptions are \utwo\ for $r>\ropt$ and \ufiveo\ for $r>2.2\ropt$), as well as $V_{obs}(r)$ determined from \hi\ and optical spectroscopy in the overlap region near \ropt. The symmetry and regularity in the inferred dynamics reflects the depth of the gravipotential wells and the similar geometries of the ionized and atomic ISM components in these massive systems. For all sample galaxies except \ueight, the \hi\ RCs extend well beyond those in the optical, probing the structure of the fast rotators out to $1.7\,\ropt \lesssim r \lesssim 3.5\,\ropt$.


 The derived RCs are very flat beyond \ropt. This is particularly the case for the highly inclined systems, where $V_{obs}(r) \simeq V_{rot}$ and projection effects are negligible. With the exception of the warp region on the approaching side of \ufiveo, $V_{obs}(r) \simeq \rm{constant}$ at $r \gtrsim 0.5\,\ropt$ for all RCs in Fig.~\ref{fast1_fig:edgeons}. For galaxies at intermediate $i$, there is an ambiguity between $i$ and the derived RC amplitude (\S\ref{fast1:inti}). In particular, the outer RCs for \none\ (Fig.~\ref{fast1_fig:intinc}a) and \ufiveone\ (Fig.~\ref{fast1_fig:intinc}c) decline on both sides when the stellar disk $i_d$ is adopted to correct for disk geometry beyond \ropt. \ufiveone\ exhibits the strongest outer RC gradient in the sample, decreasing steadily beyond \ropt\ to $V_{obs}(r_{HI})/\sin{i_d} \sim 0.75\Vrot$. However, the \hi\ channel maps of the intermediate $i$ systems are consistent with {\it both} the RCs of Fig.~\ref{fast1_fig:intinc} and flat RCs with decreasing (increasing for the approaching side of \utwo) $i$ to mimic a warp. In addition, \ufiveone\ has the lowest $i_d$ in the fast rotator sample, and a gradient $\Delta i \lesssim 1^{\circ}\, \rm{kpc}^{-1}$ is sufficient to reconcile the declining $V_{obs}(r)/\sin{i_d}$ of Fig.~\ref{fast1_fig:intinc}c with a flat RC beyond \ropt. Because of the shallow derivative of the sine function near $\sin{i}\sim90$, a gradient of this amplitude is not detectable in the outer \hi\ layers of our near edge-on systems. Indeed, since we obtain flat RCs for the highly inclined sample galaxies in which $i$ effects are negligible, and since intrinsic galaxy properties should not correlate with viewing geometry, it seems likely that the falling RCs of the intermediate-$i$ fast rotators are the result of warps rather than changes in the gravipotential wells of these systems. This suggests that {\it all} of the massive galaxies studied here have flat RCs.

\begin{figure}
\figurenum{14}
\epsscale{1.2}
\plotone{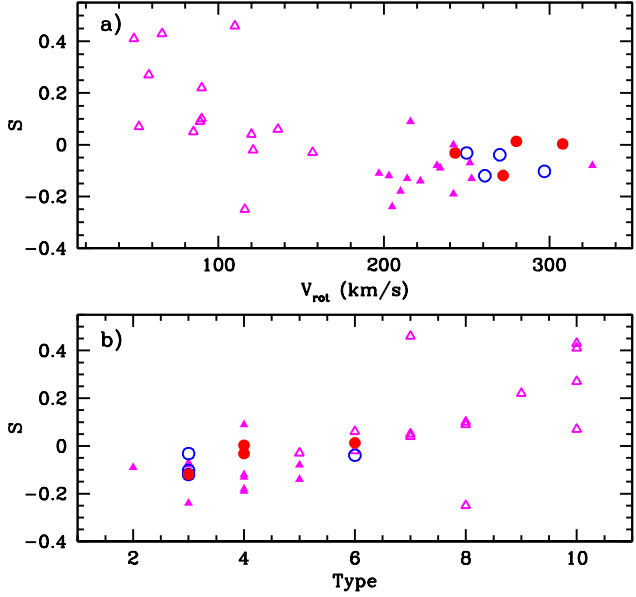} 
\caption{Logarithmic outer slope $S$ as a function of $(a)$ \Vrot\ and $(b)$ morphological type. In both panels filled circles show the highly inclined galaxies in the present sample, and open circles show those at intermediate $i$. Note that at high $i$, there is no ambiguity between $S$ and $i$. For the intermediate-$i$ galaxies, $S$ is computed using $i_d$ from Table~\ref{fast1_tab:iband}. Triangles show the sample of \hi\ RCs collected by Casertano \& van Gorkom (1991, hereafter CvG91): open symbols show systems with $V_{rot}<180\kms$, and filled symbols show systems with $V_{rot}>180\kms$. The value of $S$ decreases with increasing \Vrot\ for $\Vrot \lesssim 220\,\kms$, in accordance with the trend noted by \citetalias{cvg}; however, it plateaus in higher mass systems. The trend between $S$ and morphological type for the high-mass systems ($2 \leq \rm{type} \leq 6$) is also considerably weakened when both samples are considered.  See the electronic edition of the Journal for a color version of this figure.}
\label{fast1_fig:compare}
\end{figure}

\begin{figure}
\figurenum{15}
\epsscale{1.2}
\plotone{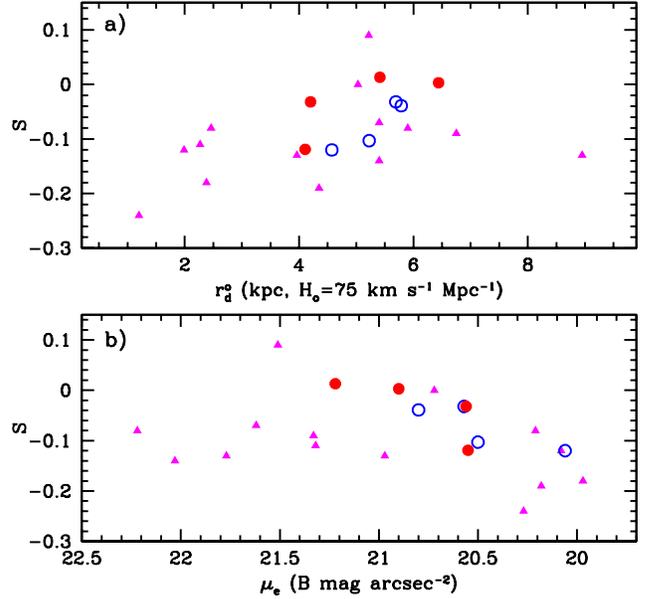} 
\caption{Logarithmic outer slope $S$ as a function of $(a)$ $r_d^o$ (computed from $H_o=75\,\kms\,\rm{Mpc}^{-1}$, as in \citetalias{cvg}) and $(b)$ $\mu_e$. Plot details are the same as in Fig.~\ref{fast1_fig:compare}. Note that \citetalias{cvg} use $B$-band photometry in their analysis, while ours is in the \Iband; we ignore this difference in $(a)$. In $(b)$, we adjust $\mu_{e,I}$ from Table~\ref{fast1_tab:iband} using $B-I$ colors derived from our data and $B_0^T$ from the RC3 where available, or else adopting $B-I=1.8$. There is some evidence that galaxies with larger $r_d^o$ have larger $S$, but the trend depends strongly on the inclusion of outliers in $S$ and $r_d^o$.  There is no clear correlation between $S$ and $\mu_e$ for massive spiral galaxies when the \citetalias{cvg} and fast rotator samples are combined. See the electronic edition of the Journal for a color version of this figure.}
\label{fast1_fig:compare2}
\end{figure}

 We quantify the outer RC shapes discussed above by computing the logarithmic slope $S=\mathrm{d}\log({V_{obs}(r)}/\sin{i_d}) /\mathrm{d}\log{r}$ for each system. We adopt the definition of \citetalias{cvg} and measure $S$ from the best fitting linear trend to the RC between $2/3\,r_{235}^o$ and the outermost RC point, where $r_{235}^o$ is the radius of the 23.5$\,\magasec$ \Iband\ isophote corrected to a face-on perspective (Table~\ref{fast1_tab:iband}; $r_{235}^o$ in the \Iband\ is roughly equivalent to $D_{25}$ in the $B$-band). We measure the slopes for the approaching and receding sides of each RC separately and compute their average weighted by the RC extent on each side to produce $S$. The values obtained are given in Table~\ref{fast1_tab:kin} and are plotted in Figs.~\ref{fast1_fig:compare} and \ref{fast1_fig:compare2}.  

 The solid circles in Figs.~\ref{fast1_fig:compare} and \ref{fast1_fig:compare2} show $S$ for the highly inclined fast rotators, and the open circles correspond to the intermediate-$i$ fast rotators.  The triangles show the sample of galaxies with extended RCs compiled by \citetalias{cvg} (their table~2): open triangles show systems with $\Vrot<180\,\kms$ (Fig.~\ref{fast1_fig:compare}), and solid triangles show systems with $\Vrot>180\,\kms$ (Figs.~\ref{fast1_fig:compare} and \ref{fast1_fig:compare2}). Note that \citetalias{cvg} compile $B$-band photometry for their sample galaxies, while ours is in the \Iband. We ignore this difference in Fig.~\ref{fast1_fig:compare2}a, since variations in $r_d$ measured in the two bands are comparable to the uncertainties on these parameters (e.g. de Jong 1996).  For ease of comparison with fig.~7 of \citetalias{cvg}, we use $H_o=75\,\kms\,Mpc^{-1}$ to convert $r_d^o$ to physical scales for the fast rotators in Fig.~\ref{fast1_fig:compare2}a. In Fig.~\ref{fast1_fig:compare2}b we adjust $\mu_{e,I}$ from Table~\ref{fast1_tab:iband} using $B-I$ colors estimated from our data and $B_T^0$ from the RC3 where available (\citealp{rc3}), or else we adopt $B-I=1.8$ (the mean for the fast rotators with $B_T^0$ estimates). Figs.~\ref{fast1_fig:compare} and ~\ref{fast1_fig:compare2} are thus analogous to figs.~6--9 of \citetalias{cvg} with the fast rotator points overplotted, although the comparisons of Fig.~\ref{fast1_fig:compare} are more reliable than those in Fig.~\ref{fast1_fig:compare2} given the different photometric bands used in the latter. 


 Figs.~\ref{fast1_fig:compare} and \ref{fast1_fig:compare2} demonstrate that the outer RC shapes for the fast rotator sample are flatter than expected from the trends in \citetalias{cvg}. This is most obvious in Fig.~\ref{fast1_fig:compare}a, where $S$ plateaus for $\Vrot \gtrsim 220\,\kms$ for {\it both} members of the present sample and those in the \citetalias{cvg} compilation: thus, the anti-correlation between $S$ and $\Vrot$ noted by \citetalias{cvg} does not extend to the high-mass end of the spiral galaxy population. Fig.~\ref{fast1_fig:compare}b also shows that the correlation between $S$ and morphological type for the high-mass systems is weakened when both samples are considered. The correlation between $r_d^o$ and $S$ Fig.~\ref{fast1_fig:compare2}a noted by \citetalias{cvg} remains, although the trend is dominated by a few outliers in $S$ and $r_d^o$. However, there is no clear correlation between $S$ and $\mu_e$ (Fig.~\ref{fast1_fig:compare2}b) when the fast rotators and the massive systems from \citetalias{cvg} are combined. We remind the reader, however, that the comparisons in Fig.~\ref{fast1_fig:compare2} are muddled by the different photometric bands employed here and in \citetalias{cvg}: an updated compilation of high-surface brightness galaxies with reliable extended RCs and homogeneous photometry would better address this issue.

 We note again here that there is an ambiguity between $i$ and $S$ for the intermediate $i$ fast rotators. We have adopted $i_d$ to compute $S$ here, and from the arguments above the open circles in Figs.~\ref{fast1_fig:compare} and \ref{fast1_fig:compare2} are likely underestimates of the true $S$ in these systems. Shifting these points towards $S=0$ in each panel does not change the above conclusions. We have also verified that varying the region in the outer RC used to compute $S$ has little impact on our results.  

 To summarize, the kinematics of the fast rotators imply that the RCs of massive, late-type spiral galaxies are flat and featureless for $r \gtrsim 0.5\,\ropt$ out to their last measured RC points, the latter in the range $1.7-3.5\,\ropt$. We thus find no convincing evidence that massive late-type galaxies have declining RCs, nor that their outer RC slopes correlate well with morphological type, scale-length or central surface brightness. If the inner regions of massive disk galaxies are baryon-dominated, the transition to dark matter domination beyond \ropt\ leaves no imprint on the resulting kinematics. The disks and halos of rapidly rotating spiral galaxies thus appear to ``conspire'' just as effectively as those in lower mass systems.

\acknowledgments
 
 Many thanks to M. P. Haynes for assistance with the SFI++ \Iband\
 images and photometry, to L. van Zee for advice regarding \hi\
 rotation curve derivations and to G. Gentile for help in implementing
 the MET method. We are also grateful to B. Catinella, C. Valotto and
 N. P. Vogt for obtaining and reducing the long-slit optical spectra,
 as well as to the many members of our group who acquired and reduced
 \Iband\ images over multiple observing runs. Thoughtful comments
 from the referee also helped improve the clarity of the text. This research has made use of the NASA/IPAC Extragalactic Database which is operated by the Jet Propulsion Laboratory, California Institute of Technology, under contract with the National Aeronautics and Space Administration. This research was partially funded by National Science Foundation grants AST-0307396 and AST-0307661.

\clearpage

\appendix

\section{Notes on Individual Systems}
\label{fast1:ind}

 In this appendix we present notes on individual systems. The channel maps are given in Figs.~\ref{fast1:1324chans}--\ref{fast1:11455chans}. All of the maps are corrected for primary beam attenuation, and the weighting adopted is the same as that used in the analysis of \S\ref{fast1:global}. For \eso, \ufiveo\ and \uone, we plot a subset of the channels near the data cube center for conciseness; otherwise, we plot the data at full spectral resolution.

\subsection{\none}
\label{fast1:1324notes}

 \none\ is a field spiral galaxy with prominent spiral structure (Fig.~\ref{fast1_fig:1324summ}a): there is a high degree of symmetry in the optical image, and no hint of a bar. The \Iband\ photometry of \citet{pat03a} is in good agreement with the values listed in Table~\ref{fast1_tab:iband}. In their compilation of bulge statistics from DSS imaging, \citet{lu00b} find a bulge shape intermediate between a boxy profile and an elliptical.  Both Giuricin \etal (2000) and Prada \etal (2003) classify \none\ as the dominant member of a galaxy pair: however, Giuricin \etal group it with the spiral KUG~0325-053 nearly a degree away, while Prada \etal identify a small companion 15\arcmin\ away. The first of these putative companions lies well outside the primary beam of our VLA observations, while the second lies at its FWHM: we are not sensitive to either. A cursory search of the \hi\ data cubes does not reveal any other companions in the vicinity of \none. 

 The channel maps for \none\ in Fig~\ref{fast1:1324chans}, as well as the global properties presented in Fig.~\ref{fast1_fig:1324summ}, are similar on the approaching and receding sides of the disk: the high degree of symmetry in \hi\ and in the optical suggests that \none\ is not interacting strongly with its neighbors. Our values of \ints, $W \sin{i}$, $V_{sys}$, and $M_{HI}$ for this system (Table~\ref{fast1_tab:properties}) are in good agreement with published single-dish data (Theureau \etal 1998; Paturel \etal 2003b). 

 There is also good agreement between the \hi\ and optical kinematics in \none\ (Fig.~\ref{fast1_fig:intinc}a).
A turnover in the \hi\ RC beyond $r\sim67$\arcsec\ ($\sim30\,\rm{kpc}$) is detected on both sides, and $V_{obs}(r)/\sin{i_d}$ decreases by $\sim30\,$\kms\ over $\sim 8\,$kpc. The turnover may signal a change in the galaxy potential in this region but is also consistent with a decrease in $i$ from $78^{\circ}$ to $62^{\circ}$ and a flat RC. We construct galaxy models with both characteristics and compare them to the observed channel maps. In both cases the models recover the distribution of \hi, although in each a change in position angle $\Delta \theta \sim 20^{\circ}$ beyond $r\sim65''$ is required to reproduce the emission from 5946 to 5925\kms\ and from 5432 to 5389\kms. The kinematics of \none\ are therefore consistent with a warp for $r\gtrsim30\,\rm{kpc}$, accompanied either by a falling RC or a systematic decrease in $i$.

\subsection{\utwo}
\label{fast1:2849notes}

 \utwo\ is an intermediate-$i$ field galaxy with a regular morphology in DSS and Two Micron All Sky Survey (2MASS) images. 
Little else other than single-dish \hi\ observations of \utwo\ are available in the literature. 

 In the channel maps for \utwo\ in Fig.~\ref{fast1:2849chans} the S/N of the \hi\ emission is low in the channels near $V_{sys}=8104\,\kms$, particularly at small $r$ where $\Sigma_{HI}(r)$ drops to $1 \,M_{\odot}\,\mathrm{pc}^{-2}$ (Fig.~\ref{fast1_fig:2849summ}d). The morphology of the \hi\ distribution is similar for the approaching and receding sides within \ropt, although there are significant differences farther out (see below). The \hi\ is more extended on the western side of the galaxy, due largely to a feature at $7866\,\kms$. Both the shape of the global profile and our estimates of \ints, $W \sin{i}$, $V_{sys}$ and $M_{HI}$ (Table~\ref{fast1_tab:properties}) are in excellent agreement with the single dish data of Paturel \etal  (2003b; see also Bottinelli \etal 1990). 

 We detect two companions to \utwo, 3.5\arcmin\ ($\sim120\,$kpc in projection) and 4.7\arcmin\ ($\sim150\,$kpc in projection) to the east and south of the main disk, respectively. There are no previously cataloged objects within 4.5\arcmin\ of either source; these systems represent new detections. The companions are only marginally resolved spatially, but channel maps (not shown) suggest that the southern parts of both galaxies are advancing and the northern parts are receding. Despite our relatively low spectral resolution, their global profiles show double peaks characteristic of inclined disks (Fig.~\ref{fast1_fig:2849summ}c): this is corroborated by the axial ratios of candidate optical counterparts that we identify on DSS images. In addition, the values of $\theta$ for the optical counterparts are roughly consistent with that expected from their \hi\ morphologies. The properties of the two companions are given in Table~\ref{fast1_tab:propertiesc}. Note that their values of $M_{HI}$ and $M_T$ resemble those of the Local Group spiral galaxy M33, which has $M_{HI}=1.8\times10^9\,\Msun$ and $M_T\sim6\times10^9\,\Msun$ (Corbelli \& Salucci 2000). 

 Although the optical RC is rising at the last measured point for \utwo, the \hi\ RC is flat for $12\,\rm{kpc} \lesssim r \lesssim 20\,\rm{kpc}$, and there is good agreement between the optical and \hi\  velocities in the overlap region. There is, however, a significant discrepancy between $V_{obs}(r)/\sin{i_d}$ on the approaching and receding sides beyond \ropt: the approaching side rises to $\sim 300\,\kms$ just beyond $r_{opt}$, but there is a smooth decrease on the receding side of $\sim2\,\kms \,$kpc. If these deviations from a flat RC are caused by warping in the outer \hi\ layer, we require $i$ to increase from $i_d=66^{\circ}$ to $i\sim 71^{\circ}$ on the approaching side beyond \ropt, and $i$ to decrease systematically from $i_d=66^{\circ}$ to $i\sim50^{\circ}$ at the last measured \hi\ point on the receding side. We model both sides of \utwo\ separately, using $V_{obs}(r)$, $\Sigma_{HI}(r)$ and $\theta_d$ as inputs. On the approaching side, simulations in which $i=i_d$ and ones in which $i$ increases to $71^{\circ}$ beyond \ropt\ both reproduce the channel map morphology shown in Fig.~\ref{fast1:2849chans}. Similarly, we construct two models of the receding side: the first adopts $i_d$ throughout, and in the second $i$ decreases as described to produce a flat RC beyond \ropt. Both simulations reproduce the observed \hi\ morphologies from 8408 to 8365\kms\ in the channel maps of Fig.~\ref{fast1:2849chans}, the first yielding a better match at $8365\,\kms$. However, neither model adequately recovers the broad, featureless emission from 8343 to 8300\kms.

 In summary, the derived RC for \utwo\ shows unusual behavior beyond \ropt; the dotted vertical line in Fig.~\ref{fast1_fig:intinc}b shows the approximate starting point of the pathologies. While we cannot distinguish between an asymmetric warp or distorted outer halo, tidal interactions between \utwo\ and its nearby, M33-like companions are the likely culprit. Deep imaging or \hi\ spectroscopy of the companions to look for irregular morphologies would confirm this hypothesis.

\subsection{ESO 563G21}
\label{fast1:563notes}

 \eso\ is a large, highly inclined galaxy in the southern sky included in the Flat Galaxy Catalog of Karachentsev \etal (1993, 1999).
The \Iband\ photometry of Paturel \etal (2003a) agrees well with the values listed in Table~\ref{fast1_tab:iband}. Using DSS images, \citet{lu00b} find an elliptical bulge in \eso, and \citet{sanchez03} detect a mild clockwise warp on both sides of the disk.  

 Because of its proximity and size, \eso\ is among the best resolved galaxies in the sample. The channel maps in Fig.~\ref{fast1:563chans} show two distinct features at either end of the main disk: one from 4921 to 4836\kms\ at $\sim8^h47^m21^s, -20^{\circ}3'30''$ and one from 4389 to 4262\kms\ at $\sim8^h47^m17^s, -19^{\circ}59'30''$. Both features are detected over multiple channels in our data, and are evident in the total \hi\ intensity map shown in  Fig.~\ref{fast1_fig:563summ}a. The southeastern extension is compact, \hi\ bright, and located well beyond $r_{opt}$ on the receding side of the disk. Qualitatively, it appears morphologically distinct from the disk emission but not kinematically different from it. The approximate starting point of the southeastern feature is indicated by a dotted vertical line in Fig~\ref{fast1_fig:edgeons}a. If associated with warp in the disk, its curvature is opposite that detected in the optical. The northeastern feature is faint and diffuse and is both morphologically and kinematically distinct from the underlying disk emission. Extraplanar emission in the vicinity of this feature is detected over 10 frequency channels yielding a characteristic width of more than $200\,\kms$, and at the sensitivity of our data it extends up to $27\,$kpc (in projection) above the midplane.

  We confirm that the spiral galaxy ESO~563G22 is a companion to \eso, with a projected separation of 5.9\arcmin\ ($\sim120\,$kpc) to the northeast of the latter. ESO~563G22 is a highly inclined spiral galaxy with a disturbed optical morphology, and 2MASS data reveals an asymmetry with more emission in the north.  We resolve the \hi\ emission of ESO~562G22 only marginally, but the southern part of the of the disk is receding and the norther part is advancing. Its basic properties are given in Table~\ref{fast1_tab:propertiesc}. While $M_{HI}$ for ESO~563G22 is comparable to that of M33, $M_T$ exceeds that of the Local Group spiral galaxy by a factor of 5 (Corbelli \& Salucci 2000). The orientation of the northeastern feature of \eso\ lies roughly along the line joining it and ESO~563G22, indicating that it may be the result of an interaction between the two systems. 

 Despite the unusual \hi\ morphology of \eso\ (Fig.~\ref{fast1_fig:563summ}a) and significant differences between $\Sigma_{HI}(r)$ for the approaching and receding sides (Fig.~\ref{fast1_fig:563summ}d), both the global profile in Fig.~\ref{fast1_fig:563summ}c and the derived RC in Fig.~\ref{fast1_fig:edgeons}a are symmetric. We find no difference between the RC derived including the northeastern and southeastern features and one derived excluding them, save for their extent on the receding side. We therefore include emission from these features in the parameters given in Table~\ref{fast1_tab:properties} and in the RC of Fig.~\ref{fast1_fig:edgeons}.  Our estimate of \ints\ is intermediate to single dish values from Haynes \etal (1999b) and Paturel \etal (2003b), and the global properties $V_{sys}$ and $W \sin{i}$ are in good agreement with published measurements. 

 Following the peak in the \hi\ intensity distribution in Fig.~\ref{fast1_fig:563summ}a in the manner described by \citet{gar02}, we measure a warp angle $\alpha = 6^{\circ}$ on the approaching side of \eso.  The approximate starting point $r_w$ of the warp is denoted by the dashed line in Fig.~\ref{fast1_fig:edgeons}a. Our model of the \eso\ data cube using $V_{obs}(r)$ and $\Sigma_{HI}(r)$ recovers the \hi\ channel maps in Fig.~\ref{fast1:563chans} if $i_d$ and $\theta_d$ are adopted throughout. These simulations do not reproduce the southeastern and northeastern features discussed above, however, and changes in $i$ and $\theta$ at large $r$ do not reconcile our models and the data in these regions. We note that the RC on the approaching side and that on the receding side within the dotted line in Fig.~\ref{fast1_fig:edgeons}a are unchanged when the \hi\ emission from the features is blanked from the data cubes. 

\subsection{\ufiveo}
\label{fast1:5010notes}

 \ufiveo\ is one of few massive isolated spiral galaxies with $cz \leq 5000$\kms\ (Karachentseva 1973; Varela \etal 2004).  We perform a cursory search for \hi-rich companions in our data that may have been missed in optical surveys, and find none. Despite its classification as an unbarred system, a strong edge-on bar is identified by \citet{lu00a} from near-IR surface brightness distributions. This is corroborated by large deviations from circular motion at small $r$ in the optical RC (Fig.~\ref{fast1_fig:edgeons}b).

 The \hi\ channel maps for \ufiveo\ in Fig~\ref{fast1:5010chans} reveal a spectacular symmetric warp. The warp on the southeastern side of the galaxy extends farther than that on the northeastern side, with emission detected up to $40\,$kpc away from the mid-plane at $3864\,$\kms.  Following the peak of the intensity distribution in Fig~\ref{fast1_fig:5010summ}a and adopting the definition of the warp angle $\alpha$ of \citet{gar02}, we measure $\alpha=23^{\circ}$ and $19^{\circ}$ on the approaching and receding sides respectively. These are extreme relative to those in the 26 edge-on systems surveyed by \citet{gar02}, being comparable to the amplitude of the largest symmetric warp that they detect and $>30\%$ greater than the largest they find in an isolated system.  Our estimate of the warp radius ($r_w\sim33\,\rm{kpc}$ on both sides) is indicated by the dashed vertical line in Fig.~\ref{fast1_fig:edgeons}b. There is a slight decrease in $\theta$ measured from our \Iband\ photometry at $20\,\rm{kpc} < r< 32\,\rm{kpc}$, representing a $\sim 1 \sigma$ deviation from $\theta_d$. If this is the signature of a warp in the optical disk, then the latter is in the opposite direction from that seen in \hi. This disconnect between stellar and \hi\ warps is common in spiral galaxies (Briggs 1990; Garc\'ia-Ruiz \etal 2002). 

 The \hi\ emission within $r_w$ is fairly symmetric about $V_{sys}$, and there is a good correspondence between $\Sigma_{HI}(r)$ on the approaching and receding sides in Fig.~\ref{fast1_fig:5010summ}d. The peak flux in Fig.~\ref{fast1_fig:5010summ}c and \ints\ in Table~\ref{fast1_tab:properties} are substantially larger than the single-dish estimates obtained from the Arecibo Radio Telescope\footnote{The Arecibo Observatory is operated by Cornell University under cooperative agreement with the NSF.} (Haynes \& Giovanelli 1984; Lewis \etal 1985; Paturel \etal 2003b; although see Bicay \& Giovanelli 1986), due to the large angular size of \ufiveo\ relative to the Arecibo $L$-band primary beam FWHM. 

 The \hi\ RC of \ufiveo\ in Fig.~\ref{fast1_fig:edgeons}b extends to $3.5\,\ropt$ when the warp regions are included, and there is good agreement between the optical and \hi\ kinematics near $r_{opt}$. There is an asymmetry in the warp region, however, where $V_{obs}(r)$ is $\sim50\,\kms$ lower on the approaching side than on the receding side.
We model \ufiveo\ using $V_{obs}(r)$ and $\Sigma_{HI}(r)$, where $\theta=\theta_d$ for $r<r_w$ and $\theta$ traces the warp ridge farther out. The simulated data cube recovers the channel map morphology when $i_d$ is adopted for both the main disk and warp emission. We experiment briefly with variations in the warp $i$ and find that $i<80^{\circ}$ in the simulations fails to reproduce the observed warp morphology in the channel maps. The RC flatness in the warp regions and the drop in RC amplitude beyond the warp radius on the approaching side therefore appear to be robust features in the kinematics of \ufiveo.

\subsection{\ufiveone}
\label{fast1:5166notes}

 \ufiveone\ is the lowest $i$ galaxy in the fast rotator sample. Its optical structure is intermediate to a flocculent and a smooth morphology (Elmegreen \& Elmegreen 1982), with complete and partial inner rings at $r=15$\arcsec\ (7.5 kpc) and $r=81\arcsec$ (41 kpc) as well as a pseudo outer ring at $r=2.4\arcmin$ (73 kpc) (de Vaucouleurs \& Buta 1980). None of these features coincides with the ``kink'' in the \Iband\ surface brightness profiles (Fig.~\ref{fast1_fig:sb}) and $R$-band (Courteau 1996) at $r\sim\ropt$ (20 kpc). 
 Ramella \etal (1995), Mahdavi \etal (2004) and Mahdavi \& Geller (2004) assign \ufiveone\ to a loose group with 7 -- 13 members; none fall within the primary beam and bandwidth of our observations. 

 The \hi\ emission of \ufiveone\ is ``clumpy'' at all frequencies in Fig.~\ref{fast1:5166chans} but overall has the expected morphology of an inclined disk. The \hi\ distribution extends farther on the advancing side of the galaxy than on the receding side (Fig.~\ref{fast1_fig:5166summ}) due to a feature from 6830 to 6787\kms. There is a sharp decrease in $\Sigma_{HI}(r)$ beyond $r_{opt}$ on both sides of the disk that coincides with the kink in the surface brightness profile. There are also enhancements in $\Sigma_{HI}(r)$ at the locations of the complete and partial optical inner rings on the approaching side. There is a hint of a warp or other distortion along the minor axis of the velocity field, but the angular resolution along that direction is relatively low (Fig.~\ref{fast1_fig:5166summ}b). The global profile in  Fig.~\ref{fast1_fig:5166summ}d is symmetric, and \ints\ is 10\%--15\% larger than previous single-dish measurements (Haynes \& Giovanelli 1984; Haynes \etal 1999b; Paturel \etal 2003b; Vogt \etal 2004). 

 The \hi\ RC for \ufiveone\ follows the upper envelope of that derived in the optical near $r_{opt}$ (Fig.~\ref{fast1_fig:intinc}c). The latter varies by $\sim80\,\kms$ in this region, likely due to spiral structure in the disk. 
There is a systematic decrease of $\sim2\,\kms\,\rm{kpc}^{-1}$ in $V_{obs}(r)/\sin{i_d}$ for $r>27\,$kpc of $\sim2\,\kms\,\rm{kpc}^{-1}$, such that the amplitude of the outermost RC point is $\sim50\kms$ lower than the value at \ropt. We construct two models of the \hi\ emission in \ufiveone\ using $V_{obs}(r)$ and $\Sigma_{HI}(r)$: the first adopts $i_d$ and $\theta_d$, and in the second we mimic a warp by gradually decreasing $i$ to $49^{\circ}$ to produce a flat RC beyond $r_{opt}$. We find slightly better agreement between the second simulation and the channel maps in  Fig.~\ref{fast1:5166chans} on the approaching side, but both models recover the receding side of the \hi\ distribution equally well. It is therefore plausible that the turnover beyond $r_{opt}$ in the HI RC in Fig.~\ref{fast1_fig:intinc}c stems from an under-estimate of $i$ in this region rather than a change in the halo potential. 

\subsection{\ueight}
\label{fast1:8220notes}

 \ueight\ is an edge-on spiral galaxy in the Coma Cluster, with a projected separation of 3.82$^{\circ}$ and a radial velocity separation of $\sim200\,\kms$ from the cluster center (Kent \& Gunn 1982).

 The channel maps of \ueight\ in Fig.~\ref{fast1:8220chans} show few deviations from the morphology of an edge-on disk. While the outer \hi\ contours from 7361 to 7360\kms\ and 6906 to 6884\kms\ bend slightly at large $r$, there is little evidence for a change in $\theta$ in the \hi\ total intensity map of Fig.~\ref{fast1_fig:8220summ}a, and we find no conclusive evidence for a warp. As in the single-dish spectrum of Freudling \etal (1988), we find a mild asymmetry in the global profile horns and in $\Sigma_{HI}(r)$. Our estimate of \ints\ is 25\% larger than published single-dish values (Giovanelli \etal 1997; Paturel \etal 2003b; Vogt \etal 2004), but there is good agreement between our results and $V_{sys}$, $W \sin{i}$ from the literature. A cursory search of the \hi\ data cubes does not reveal any companions. 

 There is excellent agreement between the optical and \hi\ kinematics in the overlap region, as well as a high degree of symmetry between the approaching and receding sides. However, the \hi\ kinematics do not extend much farther into the halo than the corresponding optical RC (Fig.~\ref{fast1_fig:edgeons}c).  We model the \ueight\ data cube using the derived kinematics and $\Sigma_{HI}(r)$, adopting $i_d$ and $\theta_d$ from Table~\ref{fast1_tab:iband}. At the angular and spectral resolution of our \hi\ data across the disk of \ueight, these simulations recover the observed frequency dependence of the \hi\ distribution.

\subsection{\uten}
\label{fast1:10469notes}

 \uten\ is an intermediate-$i$ spiral galaxy in Abell 2199 with a projected separation of 1.61$^{\circ}$ and a radial velocity separation of $\lesssim100\,\kms$ from the cluster center (Zabludoff \etal 1993). A strong $m=1$ Fourier term in $R$-band photometry implies a moderate lopsidedness in the stellar disk (Rudnick \etal 2000), and a corresponding increase in disk star formation is inferred from the strength of the Balmer absorption and the weakness of the 4000\AA\ break. The \Iband\ surface brightness distribution in Fig.~\ref{fast1_fig:sb} and the $V$-band profile of Gavazzi \etal (1994) are also unusual, with a pronounced ``dip'' for $0.2\ropt\ \lesssim 0.7\ropt$ ($3\,\rm{kpc} \lesssim r \lesssim 30\,\rm{kpc}$): this feature is not a manifestation of extinction in the system given its relatively low $i$. The shallow FIR spectral index of \uten\ between 25 and 60$\mu$m may indicate active galactic nucleus activity (Condon \etal 2002).

 The \hi\ channel maps for \uten\ in Fig.~\ref{fast1:10469chans} also show some asymmetries, particularly between the southeastern and northwestern ``wings'' of the distribution at 9044, 8912, and 8825\kms. The total intensity distribution in Fig.~\ref{fast1_fig:10469summ}a is also lopsided, and the \hi\ extends farther on the approaching side than on the receding side.  The $40\%$ difference in the peaks of the global profile in Fig.~\ref{fast1_fig:10469summ}c is less prominent in the single-dish data from Haynes \etal (1997; see also Vogt \etal 2004).
 
  We detect two companions to \uten, 5.7\arcmin\ ($215\,$kpc in projection) to the N and 4.0\arcmin\ ($150\,$kpc in projection) to the northwest of the main disk, respectively. The first is KUG~1634+392, a 16.5 mag (blue) galaxy (Garnier \etal 1996) classified as a UV-excess object by Takase \& Miyauchi-Isobe (1984). Its optical dimensions imply a low $i$; ISM turbulence may therefore be an important contributor to $W \sin{i}$, and the listed lower limit on $M_T$ is highly uncertain (Table~\ref{fast1_tab:propertiesc}).  There are no cataloged objects in the vicinity of the northwestern companion, and it is therefore a new detection. Channel maps (not shown) indicate that the northern side is advancing and the southern side is receding: this is corroborated by the morphology of a faint counterpart at the same location in our \Iband\ image. 
 Tidal interactions between \uten\ and these companions provide a natural explanation for its lopsided stellar disk and enhanced star formation rate.

 Despite the unusual stellar and gaseous morphologies of \uten, there is a high degree of symmetry in the hybrid RC (Fig.~\ref{fast1_fig:intinc}d). The optical kinematics reveal a steep inner rise, and the RC is flat at all measured $r>2\,$kpc. Although our models of the \hi\ channel maps do not reproduce the lopsidedness discussed above, there is broad general agreement between the data and simulations with $V_{obs}(r)$ and $\Sigma_{HI}(r)$ as inputs when $i_d$ and $\theta_d$ are adopted throughout the \hi\ disk.

\subsection{\uone}
\label{fast1:11455notes}

 \uone\ is an edge-on field galaxy included in Karachentseva's (1993; see also Karachentsev \etal 1999) Flat Galaxy Catalog. \citet{lu00a,lu00b} measure an elliptical bulge from surface brightness analyses of NIR and DSS images, and there is no indication of a bar from PV cuts parallel to the major axis in the NIR images. 

 The channel maps of \uone\ in Fig.~\ref{fast1:11455chans} reveal a warp on the approaching side, most prominent from 5132 to 5090\kms. There is also a low column density extension on the receding side from 5667 to 5646\kms, although $\theta$ is the same as that of the galaxy major axis. We measure a warp angle $\alpha=12^{\circ}$ on the approaching side by following the peak of the total intensity distribution (Fig.~\ref{fast1_fig:11455summ}a) as in \citet{gar02}, and the approximate starting location $r_w$ of the warp is indicated by the dashed vertical line in Fig.~\ref{fast1_fig:edgeons}d. The \hi\ distribution is symmetric about the galaxy center beyond the optical disk, and features in $\Sigma_{HI}(r)$ beyond $r \sim 70''$ (26 kpc) are well correlated.  There is also good agreement between $V_{sys}$, \ints\ and $W \sin{i}$ and single-dish estimates in the literature (Bottinelli \etal 1993; Theureau \etal 1998; Haynes \etal 1999b; Paturel \etal 2003b; Vogt \etal 2004). 

 We confirm that CGCG~341-027 is a companion to \uone, with a projected separation of 3.4\arcmin\ ($75\,$kpc in projection) to the northwest and radial velocity separation of 243\kms\ from the latter. CGCG~341-027 is a 15.7$^{th}$ (blue) mag Sb spiral, and our measurement of $V_{sys}$ is in agreement with that of Karachentsev \etal (1985). Channel maps (not shown) indicate that the southern side is advancing and the northern side is receding, and there is an asymmetry in the global profile with more gas on the receding side in Fig.~\ref{fast1_fig:11455summ}c (although the S/N is low). For comparison, $M_{HI}$ for CGCG~341-027 is half that of the Local Group spiral M33, although its rotational velocity is at least as large as in the latter (Corbelli \& Salucci 2000). The presence of a companion to \uone\ raises the possibility that the asymmetric \hi\ warp may be tidally driven.

  Although the optical RC for \uone\ is still rising at the last measured point (Fig.~\ref{fast1_fig:edgeons}), the outer \hi\ RC is flat and extends well beyond $r_{opt}$ on the advancing side. There is good agreement between the two kinematic tracers near the edge of the optical RC on both sides of the disk. We model the \hi\ emission in \uone\ using the derived RC and $\Sigma_{HI}(r)$. On the approaching side, we set $\theta=\theta_d$ for $r<r_w$ and vary $\theta$ smoothly following the warp angle $\alpha$ father out; on the receding side, we fix $\theta=\theta_d$ at all $r$. The simulated data cube recovers the channel map morphology when $i_d$ in Table~\ref{fast1_tab:iband} is adopted for both the main disk and warp emission.

\clearpage

\clearpage

\begin{landscape}
\begin{deluxetable}{ccccccccccccc}
\tablecaption{\hi\ Properties of the Fast Rotator Sample\label{fast1_tab:properties}}
\tabletypesize{\scriptsize}
\tablewidth{0pt}
\tablehead{\colhead{NGC/IC} & \colhead{\ints} & \colhead{$W \sin i$} & \colhead{$V_{sys}$} &  \colhead{$M_{HI}$} & \colhead{$r_{HI}/r_{opt}$} & \colhead{$M_T$} & \colhead{$\log(M_{HI}/{M_T})$} & \colhead{$\log(\sigma_{HI})$} & \colhead{$\log(\sigma_{T})$} & \colhead{$r_w$} & \colhead{$\alpha$}\\
 & \colhead{(Jy$\,$\kms)} & \colhead{(\kms)} & \colhead{(\kms)} & \colhead{($10^{10}\,$\Msun)} & & \colhead{($10^{11}\,$\Msun)} & & \colhead{$(\Msun \, \rm{pc}^{-2})$}  & \colhead{$(\Msun \, \rm{pc}^{-2})$} & \colhead{(kpc)} & \colhead{($^{\circ}$)}\\
\colhead{(1)} & \colhead{(2)} & \colhead{(3)} & \colhead{(4)} & \colhead{(5)} & \colhead{(6)} & \colhead{(7)} & \colhead{(8)} & \colhead{(9)} & \colhead{(10)} & \colhead{(11)} & \colhead{(12)} 
}
\startdata
\none & $10.7 \pm 0.3$ &  $583 \pm 15$ & 5671 & $1.59 \pm 0.05$ & 2.0 & $7.9 \pm 0.4$ & $-1.70\pm0.03$ & $0.52\pm0.06$ & $2.22\pm0.06$ & \nodata & \nodata\\
\utwo & $6.3 \pm 0.2$ & $506 \pm 18$ & 8124 & $2.01 \pm 0.08$ & 2.0 & $7.1 \pm 0.4$ & $-1.55\pm0.03$ & $0.57\pm0.06$ & $2.12\pm0.06$ & \nodata & \nodata \\
\eso & $27.8 \pm 0.5$ & $642 \pm 15$ & 4586 & $3.23 \pm 0.06$ & 2.5 & $14 \pm 1$ & $-1.64\pm0.03$ & $0.45\pm0.06$& $2.09\pm0.07$ & (36,0) & (6,0)\\
\ufiveo & $15.1 \pm 0.3$ & $549 \pm 23$ & $4097$ & $1.41 \pm 0.03$ & 3.3 & $8.4 \pm 0.4$ & $-1.78\pm0.02$ & $0.28\pm0.06$ & $2.06\pm0.06$ &  (33,33) & (23,19)\\
\ufiveone & $12.8 \pm 0.3$ & $441 \pm 17$ & $7009$ & $3.33 \pm 0.08$ & 2.6 & $7.7 \pm 0.7$ & $-1.36\pm0.04$ & $0.65\pm0.06$ & $2.02\pm0.07$ & \nodata & \nodata\\
\ueight & $5.2 \pm 0.2$ & $511 \pm 14$ & $7131$ & $1.32 \pm 0.06$ & 2.0 & $4.7 \pm 0.3$ & $-1.55\pm0.03$ & $0.56\pm0.07$ & $2.11\pm0.06$ & (0,0) & (0,0)\\
\uten & $5.4 \pm 0.2$ & $463 \pm 16$ & 9004 & $2.18 \pm 0.07$ & 1.8 & $5 \pm 1$ & $-1.36\pm0.09$ & $0.71\pm0.06$ & $2.1\pm0.1$ & \nodata & \nodata\\
\uone & $9.14 \pm 0.2$ & $590 \pm 17$ & $5389$ & $1.25 \pm 0.03$ & 2.0 & $8.0 \pm 0.3$ & $-1.81\pm0.02$ & $0.33\pm0.06$ & $2.13\pm0.06$ & (19,0) & (12,0)\\ 
\enddata
\tablecomments{Col. (2): Integrated line intensity. Col. (3): Velocity width at 50\% of the peak, corrected for instrumental broadening. Col. (4): Integrated profile mid-point. Col.(5): \hi\ mass, computed from eq.~\ref{fast1:b}. Col. (6): Ratio of average \hi\ radius to optical radius. Col. (7): Total dynamical mass, computed from eq.~\ref{fast1:c} (see text). Col. (9): Average \hi\ surface density $\sigma_{HI}=M_{HI}/\pi r_{HI}^2$. Col. (10):  Average total surface density $\sigma_{T}=M_{T}/\pi r_{HI}^2$. Col. (11): Measured warp radius in highly inclined systems. The first number denotes the approaching side and the second the receding side. We set $r_w=0$ when no warp is detected. Col. (12): Measured warp angle in highly inclined systems. The first number denotes the approaching side and the second the receding side.}
\end{deluxetable}
\clearpage
\end{landscape}

\end{document}